\documentclass[preprint]{aastex}

\usepackage{natbib}
\usepackage{bm}

\shorttitle{Gravitational Instability of a Porous-Dust Disk}
\shortauthors{Michikoshi \& Kokubo}

\def\vector#1{\mbox{\boldmath $#1$}}

\begin{document}
\title{
  Dynamics of Porous Dust Aggregates and Gravitational Instability of Their Disk
 }
\author{
  Shugo Michikoshi\altaffilmark{1,2}, and Eiichiro Kokubo\altaffilmark{3}
}

\email{michikos@kyoto-wu.ac.jp, and kokubo@th.nao.ac.jp }
\altaffiltext{1}{ Center for Computational Sciences, University of Tsukuba, Tsukuba, Ibaraki 305-8577, Japan }
\altaffiltext{2}{ Department for the Study of Contemporary Society, Kyoto Women's University, Imakumano, Higashiyama, Kyoto, 605-8501, Japan }
\altaffiltext{3}{ Division of Theoretical Astronomy, National Astronomical Observatory of Japan, Osawa, Mitaka, Tokyo 181-8588, Japan }

\begin{abstract}
We consider the dynamics of porous icy dust aggregates in a turbulent gas
 disk and investigate the stability of the disk.
We evaluate the random velocity of porous dust aggregates by considering
 their self-gravity, collisions, aerodynamic drag, turbulent
 stirring and scattering due to gas.
We extend our previous work by introducing the anisotropic velocity
 dispersion and the relaxation time of the random velocity.
We find the minimum
 mass solar nebular model to be gravitationally unstable if the turbulent viscosity parameter
 $\alpha$ is less than about $4 \times 10^{-3}$.
The upper limit of $\alpha$ for the onset of gravitational
 instability is derived as a function of the disk parameters.
We discuss the implications of the gravitational instability for
 planetesimal formation.
\end{abstract}

\section{Introduction}

Planetesimals are building blocks of planets
 \citep[e.g.,][]{Safronov1969, Hayashi1985}. 
The terrestrial planets and the cores of gas giants are considered to
be formed by the collisional accretion of planetesimals, which is so-called core accretion scenario
 \citep[e.g.,][]{Kokubo2012}. 
 Recently, another scenario, called the pebble accretion, has been proposed
\citep{Lambrechts2012, Lambrechts2014}. The cm-sized pebbles, loosely coupled
with gas, accrete onto planetesimals efficiently due to gas drag, which leads
to the rapid growth of gas giant cores compared to the core accretion
scenario.
It is not yet understood how dust grains grow into
 planetesimals, because this process must overcome various obstacles.
One such obstacle is the rapid radial drift of dust particles.
Because of the radial gas pressure gradient, the gas rotational
 velocity is slightly slower than the Keplerian velocity; however, 
dust particles tend to rotate with the Keplerian
 velocity.  
Thus, dust particles experience a headwind and lose their angular momentum,
 and this causes an inward radial drift
 \citep{Adachi1976, Weidenschilling1977}. 
For example, the meter-sized dust particles fall into the central star
in $10^2\mbox{--}10^3$ years before they grow into kilometer-sized planetesimals.
The meter-sized dust particles fall into the central star before they grow into kilometer-sized planetesimals. 

The gravitational instability (GI) model provides a possible solution
 to this problem \citep{Safronov1972, Goldreich1973}. 
If the gas flow is laminar, dust particles settle onto the disk midplane, due
 to the gravitational force of the central star. 
As a result, the dust layer becomes very thin.
When the dust layer density exceeds the Roche density, it becomes
 gravitationally unstable \citep{Sekiya1983, Yamoto2004, Yamoto2006}. 
Consequently, dust-rich gas clumps are formed on the dynamical timescale, and 
planetesimals form in these clumps \citep{Sekiya1983, Cuzzi2008}.
The timescale of the dust-rich gas clump formation is much faster than
 that of the radial drift, and so the GI model was considered to be a promising mechanism for 
 planetesimal formation.  

However, the turbulence in the gas stirs dust particles and prevents them
 from settling \citep{Weidenschilling1980, Youdin2007}, and even weak turbulence is sufficient to inhibit the GI.
There are various mechanisms that induce turbulence.
If the gas is sufficiently ionized, then magnetorotational instability
 causes strong turbulence \citep{Balbus1991, Sano2000}. 
Even if the gas flow is initially laminar, sedimentation of dust will
cause vertical shear instability, which leads to 
 turbulence \citep{Weidenschilling1980, Sekiya2000, Sekiya2001,
 Ishitsu2002, Ishitsu2003, Garaud2004, Michikoshi2006},  
and in the minimum mass solar nebular model this instability suppresses the GI \citep{Sekiya1998}.
Thus, the classic GI model is unlikely to fulfill its promise for explaining planetesimal formation.

Currently, the classic GI model has been replaced with various other proposed models.
One approach considers the streaming instability that is caused by the interaction between gas and
 dust \citep{Youdin2005}. During the nonlinear stage, this instability causes spontaneous particle
 clumping, which leads to the formation of gravitationally bound objects
 \citep{Johansen2007, Johansen2009, Bai2010, Bai2010a}.
This clumping is effective if the Stokes number is close to unity, and the dust-to-gas mass ratio is large
 \citep{Carrera2015, Yang2016}. 
Several mechanisms that increase the dust-to-gas ratio have been proposed; these include the inward radial drift of dust \citep{Youdin2002}, a
 migration trap in the pressure bump \citep{Haghighipour2003, Kato2012,
 Taki2016}, and the disk wind \citep{Suzuki2009, Bai2013}. 
In addition, the secular gravitational instability forms a dust-rich ring, which may result in planetesimal formation \citep{Youdin2011, Michikoshi2012, Takahashi2014, Shadmehri2016,
 Latter2017}.

Another approach considers the pairwise
 coagulation of fluffy or porous dust aggregates. 
Studies on dust growth have shown that the icy dust aggregates
 formed by pairwise coagulation can be significantly porous 
 \citep{Dominik1997, Blum2000, Wada2007, Wada2008, Wada2009,
 Suyama2008, Suyama2012}; their  
 density is typically
 $10^{-5} \, \mathrm{g}\, \mathrm{cm}^{-3}$, which is much smaller
 than that of a compact object, for which a typical density is
 $\sim 1 \, \mathrm{g}\, \mathrm{cm}^{-3}$. 
In the pairwise coagulation model, the main obstacle is the fragmentation barrier \citep{Blum2008}. 
The critical velocity of fragmentation for porous icy
 dust aggregates is larger than that for porous silicate dust
 aggregates \citep{Blum2008, Wada2009}, and
so the fragmentation barrier can be overcome relatively easily for
 icy dust.
Thus, in this paper, we focus on the evolution of icy dust aggregates.
We note that \cite{Arakawa2016} pointed out that porous silicate dust
 aggregates consisting of nanometer-sized grains can overcome the
 fragmentation barrier. 

A study of the evolution of porous icy dust aggregates showed that beyond the radial
 drift barrier,
they can grow by pairwise coagulation \citep{Okuzumi2009, Okuzumi2012}. 
The timescale of the dust growth and that of the streaming instability are
 comparable when the Stokes number is unity \citep{Okuzumi2012}.
It is not well understood which mechanism is dominant.
If the streaming instability works effectively and the Stokes number is close to unity, then gravitationally bound objects form. 
On the other hand, if pairwise coagulation is more effective than
 the streaming instability, then the dust aggregates grow and the Stokes
 number exceeds unity, and this suppresses the streaming instability.
In this paper, we postulate that dust aggregates grow sufficiently
 massive by pairwise coagulation, and the Stokes number exceeds unity. 

If large dust aggregates form by pairwise coagulation, their
 density is much smaller than that of compact planetesimals, and so it is clear that some compression mechanism is necessary.
\cite{Kataoka2013a} investigated the compression strength of icy dust
 aggregates;   
\cite{Kataoka2013} used these results to evaluate the gas
 pressure compression and the self-gravity compression and thus track the
 evolution from porous dust aggregates to compact planetesimals. 
They found that dust aggregates with mass
 $m_\mathrm{d} \gtrsim 10^{11} \mathrm{g}$ are compressed by
 self-gravity, and the density for a mass of $10^{18} \, \mathrm{g}$ reaches
 $0.1\,\mathrm{g}\,\mathrm{cm}^{-3}$. 
They concluded that planetesimals can form only by pairwise
 coagulation.  

In \cite{Michikoshi2016b} (Paper I), we considered the final stage
 of the evolution of icy dust aggregates. 
We investigated the dynamics of dust aggregates and obtained their
 random velocity.  
We found that, for a
 reasonable range of turbulence strength, the porous-dust disk becomes gravitationally unstable as
 the dust aggregates evolve through self-gravity compression. 
In Paper I, we adopted the minimum mass solar nebular model, and for simplicity,
 we assumed an isotropic velocity dispersion and an equilibrium random
 velocity. 
In this paper, we adopt a more general disk model and a more precise
 dynamical model and confirm the results of Paper I.
Here, we consider an anisotropic velocity dispersion, that is, 
 the evolution of the eccentricity and that of the inclination are calculated
 separately. 
The relaxation time of the random velocity is also taken into account,  
 since in some cases, it is comparable to the timescale of dust growth.

The remainder of the paper is organized as follows.
In Section \ref{sec:method}, we present the model of the protoplanetary
 disk and dust aggregates. 
In Section \ref{sec:dyncamics}, we explain the porous dust dynamics model, and in Section \ref{sec:params}, we calculate the equilibrium eccentricity and
 inclination, and investigate the stability of the porous-dust disk. 
We discuss the general properties of the disk independent from the evolution of the dust.
In Section \ref{sec:onset}, we investigate the disk stability as it undergoes
compression by self-gravity, and we derive a 
condition for the onset of gravitational instability. 
Section \ref{sec:summary} is devoted to a summary and discussion of our results.

\section{Model \label{sec:method}}

\subsection{Protoplanetary Disk}
  
We adopt the following surface densities of gas and dust \citep{Weidenschilling1977a, Hayashi1981}: 
\begin{eqnarray}
 \Sigma_\mathrm{g} & = &
  1700 f_\mathrm{g} \left(\frac{a}{\mathrm{1\,AU}}\right)^{-\beta_\mathrm{g}} \,
 \mathrm{g}\, \mathrm{cm}^{-2}, \\
 \Sigma_\mathrm{d} & = & \gamma \Sigma_\mathrm{g},
\end{eqnarray}
 where $a$ is the distance from the central star, $\beta_\mathrm{g}$ is
 the power-law index, $f_\mathrm{g}$ is the ratio in the minimum-mass
 solar nebula (MMSN) model, and $\gamma$ is the dust-to-gas mass ratio.  
As the fiducial model, we consider the MMSN model
 ($\beta_\mathrm{g}=3/2$ and $\gamma=0.018$) beyond the snowline
 \citep{Hayashi1981, Hayashi1985}. 
We adopt the temperature profile
\begin{equation}
  T = T_1 \left( \frac{a}{1\,\mathrm{AU}} \right)^{-\beta_\mathrm{t}} \,\mathrm{K},
\end{equation}
 where $T_1$ is the temperature at $1\,\mathrm{AU}$, and
 $\beta_\mathrm{t}$ is the power-law index. 
In the fiducial model, we adopt $T_1=120$ and $\beta_\mathrm{t}=3/7$
 \citep{Chiang2010}.
The isothermal sound velocity is calculated from the temperature as
 $c_\mathrm{s}=\sqrt{k_\mathrm{B}T/m_\mathrm{g}}$, where $k_\mathrm{B}$ 
 is the Boltzmann constant, and
 $m_\mathrm{g} = 3.9 \times 10^{-24} \, \mathrm{g}$ is the mean molecular mass. 
The gas density at the disk midplane is
 $\rho_\mathrm{g} = \Sigma_\mathrm{g}/(\sqrt{2 \pi} c_\mathrm{s}/\Omega)$,
 where $\Omega = \sqrt{G M_*/a^3}$ is the Keplerian angular frequency,
 and $M_*$ is the mass of the central star.
Throughout this paper, we adopt $M_*=M_\odot$, where $M_\odot$ is the
 solar mass. 
The mean free path of gas molecules is
 $l = m_\mathrm{g} / \sigma_\mathrm{g} \rho_\mathrm{g}$, where
 $\sigma_\mathrm{g} = 2 \times 10^{-15}\, \mathrm{cm}^2$ is the
 collisional cross-section of gas molecules.
The nondimensional radial pressure gradient is given as \citep{Nakagawa1986}
\begin{equation}
 \eta = -\frac{1}{2} \left( \frac{c_\mathrm{s}}{a \Omega} \right)^2
 \frac{\partial \log (\rho_\mathrm{g} c_\mathrm{s}^2)}{\partial \log a} =
 (2.39 \beta_\mathrm{g} + 1.20 \beta_\mathrm{t}+ 3.59)\times10^{-4}
 \left(\frac{T_1}{120 } \right)
 \left(\frac{a}{1\,\mathrm{AU}} \right)^{1-\beta_\mathrm{t}}.
 \label{eq:etamodel}
\end{equation}
For the fiducial model ($T_1 = 120$, $\beta_\mathrm{g} = 3/2$ and
 $\beta_\mathrm{t} = 3/7$), $\eta$ at 1 AU is $0.77 \times 10^{-3}$.

\subsection{Dust Aggregate}
\subsubsection{Physical Properties}
 
For simplicity, we will consider equal-mass dust aggregates with
 mass $m_\mathrm{d}$. 
We assume that a dust aggregate consists of $N$ monomers, where $N = m_\mathrm{d}/m_0$. 
The monomer mass, density, and radius are $m_0$, $\rho_0$, and $r_0$,
 respectively, which satisfy $m_0 = \frac{4\pi}{3} \rho_0 r_0^3$. 
The gyration radius of a dust aggregate is given as
 \citep{Mukai1992, Wada2008} 
\begin{equation}
  r_\mathrm{g} = \sqrt{\frac{1}{N} \sum_k ( \vector{x}_k - \vector{X})^2},
\end{equation}
 where $\vector{x}_k$ is the position of monomer $k$, and
 $\vector{X} = \sum_k \vector{x}_k /N$ is the position of the center of mass. 
The characteristic radius is defined by \citep{Mukai1992}
\begin{equation}
 r_\mathrm{c} = \sqrt{\frac{5}{3}} r_\mathrm{g}.
\end{equation}
If the dust aggregates grow by ballistic cluster-particle aggregation
 (BCPA), each dust aggregate can be considered as a uniform sphere, where 
$r_\mathrm{c}$ corresponds to the physical radius.
On the other hand, if the dust aggregates grow by ballistic
 cluster-cluster aggregation (BCCA), the dust aggregates are 
 inhomogeneous with a fractal dimension of less than 3, and
 $r_\mathrm{c}$ corresponds to the maximum distance from the approximate center of
 mass \citep{Okuzumi2009}.
We define the collisional cross-section by using $r_\mathrm{c}$.
The characteristic volume and the mean internal density are calculated
 from $r_\mathrm{c}$ as follows: 
\begin{equation}
 V_\mathrm{c} = \frac{4\pi r_\mathrm{c}^3}{3},
\end{equation}
\begin{equation}
 \rho_\mathrm{int} = \frac{m_\mathrm{d}}{V_c}.
\end{equation}
The density of a dust aggregate with fractal dimension $D$ is
\begin{equation}
 \rho_\mathrm{int} = \left(\frac{m_\mathrm{d}}{m_0} \right)^{1-3/D} \rho_0.
 \label{eq:BCCA}
\end{equation}
The fractal dimension of a BCCA cluster is $D\simeq 1.9$
 \citep{Mukai1992, Okuzumi2009}, and  
thus the density of a BCCA cluster is
 $\rho_\mathrm{BCCA} = (m_\mathrm{d}/m_0 )^{-0.58} \rho_0$. 

The interaction of dust aggregates with gas is characterized by the
 projected area $A$.  
 Generally, the projected area $A$ can be smaller than $\pi r_\mathrm{c}^2$.
\cite{Okuzumi2009} obtained the projected area formula for both BCPA and BCCA clusters
\begin{equation}
 A = \left(\frac{1}{A_\mathrm{BCCA}} + \frac{1}{\pi r_\mathrm{c}^2} -
     \frac{1}{\pi r_\mathrm{c,BCCA}^2}\right)^{-1}, 
\end{equation}
 where $A_\mathrm{BCCA}$ is the projected area for the BCCA cluster
 \citep{Minato2006} 
\begin{equation}
 \frac{A_\mathrm{BCCA}}{\pi r_0^2} \simeq
 \left\{ \begin{array}{ll}
	12.5 N^{0.685} \exp( -2.53/N^{0.0920}) & (N<16) \\
	(0.352 N + 0.566 N^{0.862}), & (N\geq 16)
 \end{array} \right.,
\end{equation}
 and $r_\mathrm{c,BCCA}$ is the characteristic radius of the
 corresponding BCCA cluster, 
\begin{equation}
 \pi r_\mathrm{c,BCCA}^2 \simeq N^{2/D} \pi r_0^2.
\end{equation}
We define the area-equivalent radius from $A$ as
\begin{equation}
 r_\mathrm{A} = \sqrt{\frac{A}{\pi}}.
\end{equation}

\begin{figure}
\plotone{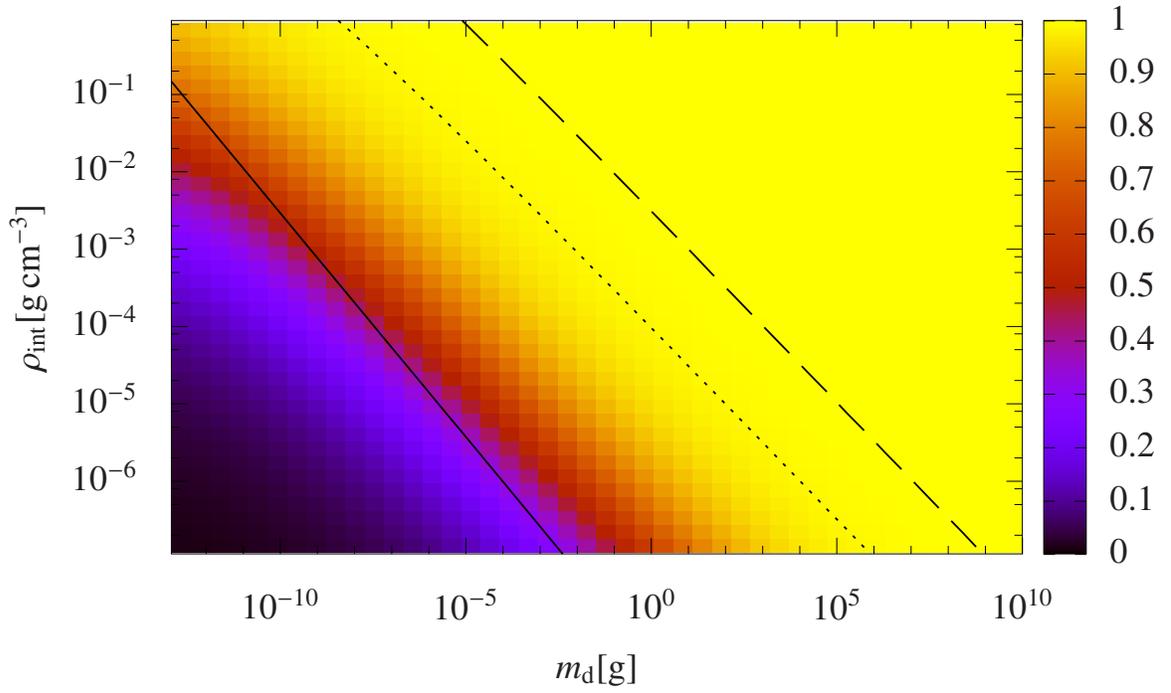}
\caption{
Area-equivalent radius normalized by the characteristic radius
 $r_\mathrm{A}/r_\mathrm{c}$ on the $m_\mathrm{d}$-$\rho_\mathrm{int}$ plane.
The dashed and dotted lines show $r_\mathrm{A}/r_\mathrm{c} = 0.999$ and
 $0.99$, respectively.
The solid line shows the BCCA cluster density $\rho_\mathrm{BCCA}$.
\label{fig:ra_rc}
}
\end{figure}

The area-equivalent radius is shown in Figure \ref{fig:ra_rc}.
If the density is sufficiently larger than the BCCA cluster density
 $\rho_\mathrm{BCCA}$, $r_\mathrm{A}$ is approximated by $r_\mathrm{c}$. 
Since we investigate the evolution due to self-gravity compression, we
 consider the following range of parameters: $m_\mathrm{d} > 10^{10} \mathrm{g}$
 and $\rho_\mathrm{int} > 10^{-5} \mathrm{g}\, \mathrm{cm}^{-3}$
 \citep{Kataoka2013}. 
In this parameter region, the density is much larger than
 $\rho_\mathrm{BCCA}$, and the corresponding fractal dimension is larger
 than $2$.
Thus, throughout this paper, we assume $r_\mathrm{A} \simeq r_\mathrm{c}$.
In short, we consider a dust aggregate to be a sphere with radius
 $r_\mathrm{c}$, which has the usual mass-radius relation
 $m_\mathrm{d} = (4 \pi/3)\rho_\mathrm{int} r_\mathrm{c}^3$. 

\subsubsection{Evolution \label{sec:evotrack}}

We consider the evolution of dust due to coagulation and self-gravity
 compression \citep{Kataoka2013}. 
The compressive strength of a porous dust aggregate is \citep{Kataoka2013a}
\begin{equation}
 P_{\mathrm{comp}} \simeq 
 \frac{E_\mathrm{roll}}{r_0^3} \left(\frac{\rho_\mathrm{int}}{\rho_0} \right)^3,
\end{equation}
 where $E_\mathrm{roll}$ is the rolling energy \citep{Dominik1997, Wada2007}.
Equilibrium is obtained when
 $P_\mathrm{comp} = P$ \citep{Kataoka2013}, and 
from this we can calculate the equilibrium density.
The self-gravity pressure is
\begin{equation}
 P_{\mathrm{grav}} \simeq \frac{Gm_\mathrm{d}^2}{\pi r_\mathrm{c}^4},
\end{equation}
and thus we obtain the equilibrium density \citep{Kataoka2013}
\begin{equation}
 \rho_\mathrm{eq} \simeq
 1.58 \left(\frac{r_0^3 \rho_0^3 G}{E_\mathrm{roll}}\right)^{3/5}
 m_\mathrm{d}^{2/5}.
 \label{eq:evotrack}
\end{equation}
As the mass of the dust aggregate increases, its self-gravity pressure increases;
 this compresses it and increases its equilibrium density. 
 In the fiducial model, we adopt $E_\mathrm{roll}=4.74\times 10^{-9}\,\mathrm{erg}$, $\rho_0=1.0\,\mathrm{g}\,\mathrm{cm}^{-3}$, and $r_0 = 0.1\,\mu \mathrm{m}$.

\section{Dynamics \label{sec:dyncamics}}
We define $\sigma_e$ and $\sigma_i$ as the root mean squares of the eccentricity and inclination, respectively, of the dust aggregates.
We developed a model to calculate the evolution of $\sigma_e$ and $\sigma_i$, taking into account gravitational scattering, collisions among dust aggregates, and interactions with gas.

\subsection{Gravitational Scattering}

Gravitational scattering among dust aggregates results in an increase in both $\sigma_e$ and $\sigma_i$
 \citep{Ida1990, Stewart2000}.
In Paper I, we adopted the simple stirring rate formula described by the
 Chandrasekhar relaxation time \citep{Ida1990}, which is valid in the
 dispersion-dominated regime.
\cite{Ohtsuki2002} examined the evolution of $\sigma_e$ and $\sigma_i$ using
 three-body integration, and they derived a semianalytic formula that is
 valid in both the dispersion-dominated and the shear-dominated regime.
In this paper, we adopt their stirring rates:
\begin{equation}
 \left( \frac{\mathrm{d} \sigma_e^2}{\mathrm{d}t} \right)_\mathrm{grav} =
 \frac{a^2 \Omega h^4 \Sigma_\mathrm{d}}{4m_\mathrm{d}} P_\mathrm{VS},
\end{equation}
 and
\begin{equation}
 \left( \frac{\mathrm{d}\sigma_i^2}{\mathrm{d}t} \right)_\mathrm{grav} =
 \frac{a^2 \Omega h^4 \Sigma_\mathrm{d}}{4m_\mathrm{d}} Q_\mathrm{VS},
\end{equation}
 where $h$ is the reduced Hill radius
 $h=(2m_\mathrm{d}/3M_*)^{1/3}$ and 
\begin{equation}
 P_\mathrm{VS} =
  73 \frac{\log(10 \Lambda^2/\tilde \sigma_{e,\mathrm{r}}^2 + 1) }{ 10 \Lambda^2/\tilde \sigma_{e,\mathrm{r}}^2}  + \log(\Lambda^2+1) \frac{72}{\pi \tilde \sigma_{e,\mathrm{r}} \tilde \sigma_{i,\mathrm{r}}} \int_0^1 \frac{5 K_\lambda - 12 (1- \lambda^2)E_\lambda/(1+3\lambda^2)}{\beta+(1/\beta-\beta)\lambda^2} \mathrm{d} \lambda,
\end{equation}
\begin{equation}
 Q_\mathrm{VS} =
  (4 \tilde \sigma_{i,\mathrm{r}}^2 + 0.2 \tilde \sigma_{e,\mathrm{r}}^3\tilde \sigma_{i,\mathrm{r}})\frac{\log(10 \Lambda^2 \tilde \sigma_{e,\mathrm{r}} + 1)}{ 10 \Lambda^2 \tilde \sigma_{e,\mathrm{r}}} + \log(\Lambda^2+1) \frac{72}{\pi \tilde \sigma_{e,\mathrm{r}} \tilde \sigma_{i,\mathrm{r}}} \int_0^1 \frac{K_\lambda - 12 \lambda^2E_\lambda/(1+3\lambda^2)}{\beta+(1/\beta-\beta)\lambda^2} \mathrm{d} \lambda,
\end{equation}
 where $\beta = \tilde \sigma_{i,\mathrm{r}}/\tilde \sigma_{e,\mathrm{r}}$,
 $K_\lambda=K(\sqrt{3(1-\lambda^2)}/2)$,
 $E_\lambda=E(\sqrt{3(1-\lambda^2)}/2)$, and
 $\Lambda = \tilde \sigma_{i,\mathrm{r}} (\tilde \sigma_{e,\mathrm{r}}^2 + \tilde \sigma_{i,\mathrm{r}}^2)/12$.
 The first and second terms correspond to the low-velocity and high-velocity limits, respectively.
The reduced relative eccentricity $\tilde \sigma_{e,\mathrm{r}}$ and the inclination
 $\tilde \sigma_{i,\mathrm{r}}$ are
 $\tilde \sigma_{e,\mathrm{r}} = \sqrt{2}\sigma_e/h$ and
 $\tilde \sigma_{i,\mathrm{r}} = \sqrt{2}\sigma_i/h$, respectively \citep{Nakazawa1989}.
The functions $K(k)=\int_0^{\pi/2} (1-k^2 \sin^2 \theta)^{-1/2} \mathrm{d} \theta$ and $E(k)=\int_0^{\pi/2} (1-k^2 \sin^2 \theta)^{1/2} \mathrm{d} \theta$ are the complete elliptic integrals of the first and second kind, respectively, where $k$ is the modulus.

\subsection{Collisions}
We assume that the average change in $\sigma_e$ for a dust collision is
 $\sigma_e^2 \to C_\mathrm{col} \sigma_e^2$.
We adopt $C_\mathrm{col} = 1/2$, which corresponds to perfect
 accretion \citep{Inaba2001}.
We discuss the effect of the imperfect accretion on the dust aggregate growth in Section \ref{sec:noneq}.
Using the nondimensional collision rate $P_\mathrm{col}$
 \citep{Nakazawa1989}, the evolution equations for $\sigma_e$ and $\sigma_i$ are
 written as
\begin{equation}
 \left( \frac{\mathrm{d} \sigma_e^2}{\mathrm{d}t} \right)_\mathrm{col} =
 -C_\mathrm{col} P_\mathrm{col} h^2 a^2
  \frac{ \Sigma_\mathrm{d}}{m_\mathrm{d}} \Omega \sigma_e^2,
  \label{eq:damp_col}
\end{equation}
\begin{equation}
 \left( \frac{\mathrm{d}\sigma_i^2}{\mathrm{d}t} \right)_\mathrm{col} =
 -C_\mathrm{col} P_\mathrm{col} h^2 a^2
  \frac{ \Sigma_\mathrm{d}}{m_\mathrm{d}} \Omega \sigma_i^2.
\end{equation}
For the low-velocity regime, where $\tilde \sigma_e, \tilde \sigma_i < 0.2$,
 $P_\mathrm{col}$ is independent of $\tilde \sigma_e$ and $\tilde \sigma_i$ \citep{Ida1989, Inaba2001}, and
\begin{equation}
 P_\mathrm{col} \simeq  P_\mathrm{low} = 11.3 \sqrt{\tilde r},
\end{equation}
 where $\tilde \sigma_e = \sigma_e/h $ and $\tilde \sigma_i = \sigma_i/h $ are the reduced eccentricity and inclination, respectively, and $\tilde r = 2r_\mathrm{c}/ha$.
For the medium-velocity regime, where $0.2 < \tilde \sigma_e, \tilde \sigma_i < 2$ \citep{Ida1989, Inaba2001},
 $P_\mathrm{col}$ depends on $\tilde \sigma_{i,\mathrm{r}}$:
\begin{equation}
 P_\mathrm{col} \simeq  P_\mathrm{med} =
  \frac{\tilde r^2}{4 \pi \tilde \sigma_{i,\mathrm{r}}}
  \left(17.3 + \frac{232}{\tilde r}\right).
\end{equation}
For the high-velocity regime, where $2 < \tilde \sigma_e, \tilde \sigma_i $, $P_\mathrm{col}$
 is \citep{Greenzweig1992}
\begin{equation}
 P_\mathrm{col} \simeq P_\mathrm{high} =
 \frac{\tilde r^2}{2 \pi} \left( 8 \int_0^1 \mathrm{d} \lambda
  \frac{\beta^2 E_\lambda }{(\beta^2 + (1-\beta^2)\lambda^2)^2} +
  \frac{48}{\tilde r \tilde \sigma_{e,\mathrm{r}}^2} \int_0^1 \mathrm{d} \lambda
  \frac{K_\lambda }{\beta^2 + (1-\beta^2)\lambda^2} \right),
  \label{eq:pcoleq}
\end{equation}
 where the second term indicates the effect of gravitational focusing.
\cite{Inaba2001} proposed the following formula for the general nondimensional collision rate:
\begin{equation}
 P_\mathrm{col} =
 \mathrm{min}
  (P_\mathrm{med}, (P_\mathrm{high}^{-2} + P_\mathrm{low}^{-2})^{-1/2}).
\label{eq:inaba}
\end{equation}
We set $P_\mathrm{col} = P_\mathrm{high}$ for a large random velocity, such as $\tilde \sigma_i > 10$. 
Note that this formula does not take into account the gas drag effect, which may alter the collision rate.  
In the present model, when the gas drag is strong, turbulent stirring causes the random velocity to be high.  In this case, the collision rate formula is described by the geometrical cross-section. 

\subsection{Gas Effects}
\subsubsection{Gas Drag}

Because of the hydrodynamic gas drag, $\sigma_e$ and $\sigma_i$ decrease with time as follows
 \citep{Adachi1976, Inaba2001}:
\begin{equation}
 \left(\frac{\mathrm{d} \sigma_e^2}{\mathrm{d}t}\right)_\mathrm{gas,drag} =
 - \frac{2}{t_{\mathrm{s},e}} \sigma_e^2,
  \label{eq:eeq}
\end{equation}
\begin{equation}
 \left(\frac{\mathrm{d} \sigma_i^2}{\mathrm{d}t}\right)_\mathrm{gas,drag} =
 - \frac{2}{t_{\mathrm{s},i}} \sigma_i^2,
  \label{eq:ieq}
\end{equation}
 where $t_{\mathrm{s},e}$ and $t_{\mathrm{s},i}$ are the damping
 timescales of $\sigma_e$ and $\sigma_i$:
\begin{eqnarray}
  t_{\mathrm{s},e} &=&
  \frac{2 m_\mathrm{d}}{\pi C_\mathrm{D} r_\mathrm{c}^2 \rho_\mathrm{g} v_\mathrm{K} \left(\displaystyle{\frac{9 E^2}{4 \pi} \sigma_e^2} + \displaystyle{ \frac{1}{\pi} \sigma_i^2 } + \displaystyle{ \frac{9}{4} \eta^2} \right)^{1/2}}, \label{eq:te} \\
  t_{\mathrm{s},i} &=&
  \frac{4 m_\mathrm{d}}{\pi C_\mathrm{D} r_\mathrm{c}^2 \rho_\mathrm{g} v_\mathrm{K} \left(\displaystyle{\frac{ E^2}{\pi} \sigma_e^2} + \displaystyle{\frac{4}{\pi} \sigma_i^2} + \eta^2 \right)^{1/2}}, \label{eq:ti}
\end{eqnarray}
 where $E=E(\sqrt{3/4})$ is the elliptic integral of the second kind of argument $\sqrt{3/4}$, $v_\mathrm{K} = a \Omega$ is the Keplerian velocity, and $C_\mathrm{D}$ is the nondimensional drag coefficient. We also adopted the corrections discussed by \cite{Kary1993} and \cite{Inaba2001}.

The relative velocity between gas and dust is \citep{Adachi1976}
\begin{equation}
  u \simeq \left(v_\mathrm{ran}^2 + \eta^2 v_\mathrm{K}^2 \right)^{1/2},
  \label{eq:relative}
\end{equation}
where 
$v_\mathrm{ran} = \displaystyle{\left(\frac{5}{8}\sigma_e^2+ \frac{1}{2}\sigma_i^2\right)^{1/2}} v_\mathrm{K}$ is the random velocity.
Using the relative velocity given by Equation (\ref{eq:relative}), we define the stopping time
\begin{equation}
t_\mathrm{s}  = \frac{2 m_\mathrm{d}}{\pi C_\mathrm{D} r_\mathrm{c}^2 \rho_\mathrm{g} u },
  \label{eq:ts1}
\end{equation}
which is nearly equal to $t_{\mathrm{s},e}$ and $t_{\mathrm{s},i}$.

The gas drag law changes with $r_\mathrm{c}$ \citep[e.g.,][]{Adachi1976}.
If $r_\mathrm{c} \gtrsim l$, we use the Stokes drag or the Newton drag.
For a low Reynolds number ($\mathrm{Re} \ll 10^3$), the drag
 coefficient is approximated by $C_\mathrm{D} \simeq 24/\mathrm{Re}$
 (Stokes drag), where $\mathrm{Re} = 2 r_\mathrm{c} u/\nu$. 
The viscosity $\nu$ is given by $\nu = v_\mathrm{th} l/2$, where
 $v_\mathrm{th} = \sqrt{8/\pi} c_\mathrm{s}$ is the thermal velocity.
For a high Reynolds number ($10^3 < \mathrm{Re} < 2 \times 10^5$), the
 drag coefficient is almost constant,
 $C_\mathrm{D} \simeq 0.4 \mbox{--} 0.5$ (Newton drag).
If $r_\mathrm{c} \lesssim l$, we use the Epstein drag.
Thus, we adopt the drag coefficient formula \citep{Brown2003}
\begin{equation}
 C_\mathrm{D} =
 \left\{
  \begin{array}{ll}
  \displaystyle {\frac{8 v_\mathrm{th}}{3 u}}  & (r_\mathrm{c} < 9l/4) \\
  \displaystyle {\frac{0.407}{1+8710/ \mathrm{Re}}+ \frac{24}{\mathrm{Re}}(1+0.150 \mathrm{Re}^{0.681}) } & (r_\mathrm{c} > 9l/4)
  \end{array} \right..
  \label{eq:Cd}
\end{equation}

\subsubsection{Turbulent Stirring}

The random velocity of dust aggregates increases due to the gas drag from the
 turbulent velocity field as \citep{Michikoshi2016b}
\begin{equation}
 \frac{\mathrm{d}v_\mathrm{ran}^2}{\mathrm{d}t} =
  \frac{2 \tau_\mathrm{e} v_\mathrm{t}^2 \Omega }{S(\tau_\mathrm{e}+S)},
  \label{eq:turbstir}
\end{equation}
 where $S = \Omega t_\mathrm{s}$ is the Stokes number, $v_\mathrm{t}=\sqrt{\alpha} c_\mathrm{s}$ is the magnitude of the turbulent velocity, $\alpha$ is the dimensionless turbulence strength \citep{Cuzzi2001}, and $\tau_\mathrm{e} \Omega^{-1}$ is the eddy turnover time.
In the fiducial model, we adopt $\tau_\mathrm{e} = 1$ \citep{Youdin2011}.
For the isotropic turbulent velocity, the heating rate of
 $\sigma_e$ would be twice as large as that of $\sigma_i$, and thus we adopt the following formulae \citep{Kobayashi2016}
\begin{equation}
 \left(\frac{\mathrm{d}\sigma_e^2}{\mathrm{d}t}\right)_\mathrm{turb,stir} =
  \frac{4 \tau_\mathrm{e} v_\mathrm{t}^2 \Omega }{3 v_\mathrm{K}^2 S(\tau_\mathrm{e}+S)},
 \label{eq:turbheat_e}
\end{equation}
\begin{equation}
 \left(\frac{\mathrm{d}\sigma_i^2}{\mathrm{d}t}\right)_\mathrm{turb,stir} =
  \frac{2 \tau_\mathrm{e} v_\mathrm{t}^2 \Omega }{3 v_\mathrm{K}^2 S(\tau_\mathrm{e}+S)}.
\end{equation}

\subsubsection{Turbulent Scattering}

The fluctuations in gas density due to turbulence results in gravitational scattering of the
 dust aggregates.
\cite{Okuzumi2013} derived the stirring rate of $\sigma_e$ due to this effect:
\begin{equation}
 \left(\frac{\mathrm{d}\sigma_e^2}{\mathrm{d}t}\right)_\mathrm{turb,grav} =
 C_\mathrm{turb} \alpha
  \left( \frac{\Sigma_\mathrm{g} a^2}{M_*} \right)^2 \Omega,
\label{eq:okuzumi_formula2013}
\end{equation}
 where $C_\mathrm{turb}$ is a nondimensional coefficient.
From the semianalytical discussion, the nondimensional coefficient was
 obtained as \citep{Okuzumi2011b, Gressel2012, Okuzumi2013}
\begin{equation}
  C_\mathrm{turb} = \frac{0.94 \mathcal{L} }{(1 + 4.5 H_\mathrm{res,0}/H)^2},
\end{equation}
 where $H$ is the gas scale height, $H_\mathrm{res,0}$ is the 
 vertical half-width of the dead zone for the magneto-rotational instability,
 and $\mathcal{L}$ is a nondimensional saturation limiter that is less than or equal to unity.
For simplicity, we set $\mathcal{L} = 1$.
In the fiducial model, we adopt $H_\mathrm{res,0} = H$, which leads to
 $C_\mathrm{turb} = 3.1 \times 10^{-2}$.
In Section \ref{sec:onset}, we discuss the effect of  $C_\mathrm{turb}$.
The stirring rate of $\sigma_i$ is
\begin{equation}
 \left(\frac{\mathrm{d}\sigma_i^2}{\mathrm{d}t}\right)_\mathrm{turb,grav} =
 \epsilon_i^2 \frac{\mathrm{d}\sigma_e^2}{\mathrm{d}t},
 \label{eq:heat_i_gravsca}
\end{equation}
 where $\epsilon_i$ is a nondimensional coefficient and is smaller than
 unity \citep{Yang2012}.
We adopt $\epsilon_i =0.1$ \citep{Kobayashi2016}.

\subsection{Equilibrium Random Velocity}

Considering the above processes, we obtain the evolution equations for $\sigma_e$ and
 $\sigma_i$:
\begin{equation}
 \frac{\mathrm{d}\sigma_e^2}{\mathrm{d}t} =
 \left(\frac{\mathrm{d}\sigma_e^2}{\mathrm{d}t} \right)_\mathrm{grav} +
 \left(\frac{\mathrm{d}\sigma_e^2}{\mathrm{d}t} \right)_\mathrm{col} +
 \left(\frac{\mathrm{d}\sigma_e^2}{\mathrm{d}t} \right)_\mathrm{gas,drag} +
 \left(\frac{\mathrm{d}\sigma_e^2}{\mathrm{d}t} \right)_\mathrm{turb,stir} +
 \left(\frac{\mathrm{d}\sigma_e^2}{\mathrm{d}t} \right)_\mathrm{turb,grav},
\label{eq:evoe}
\end{equation}
\begin{equation}
 \frac{\mathrm{d}\sigma_i^2}{\mathrm{d}t} =
 \left(\frac{\mathrm{d}\sigma_i^2}{\mathrm{d}t} \right)_\mathrm{grav} +
 \left(\frac{\mathrm{d}\sigma_i^2}{\mathrm{d}t} \right)_\mathrm{col} +
 \left(\frac{\mathrm{d}\sigma_i^2}{\mathrm{d}t} \right)_\mathrm{gas,drag} +
 \left(\frac{\mathrm{d}\sigma_i^2}{\mathrm{d}t} \right)_\mathrm{turb,stir} +
 \left(\frac{\mathrm{d}\sigma_i^2}{\mathrm{d}t} \right)_\mathrm{turb,grav}.
\label{eq:evoi}
\end{equation}
If the relaxation of $\sigma_e$ and $\sigma_i$ is sufficiently fast, we can obtain the
 equilibrium values of $\sigma_e$ and $\sigma_i$ from $\mathrm{d}\sigma_e^2/\mathrm{d}t=0$ and
 $\mathrm{d}\sigma_i^2/\mathrm{d}t=0$.
In Section \ref{sec:params}, we discuss the GI using the equilibrium values of
 $\sigma_e$ and $\sigma_i$; however, we note that the validity of this treatment is not trivial.
The nonequilibrium effect is discussed in Section \ref{sec:onset}.

\subsection{Condition for Gravitational Instability}

For a dust layer that is perfectly coupled with an incompressible
 fluid, the Roche density $\rho_\mathrm{R}$ is often used for the GI
 condition \citep{Sekiya1983, Yamoto2004}.
In this case, a buckling mode develops, and dust-rich gas clumps form.
Planetesimals may form in these dense clumps \citep{Sekiya1983, Cuzzi2008}.
However, in this paper, we focus on large dust aggregates that have a large
 Stokes number.
In other words, dust aggregates can move relative to the gas, and dust can
 collapse through the gas, where the Roche criterion may not be applicable.
In this case, Toomre's $Q$ better describes the GI condition
 \citep{Toomre1964}.
We approximate the dust layer as a fluid with a finite velocity
 dispersion and calculate $Q$ as
\begin{equation}
 Q = \frac{v_x \Omega}{C_\mathrm{T} G \Sigma_\mathrm{d}},
\label{eq:qvalue}
\end{equation}
 where $v_x = (v_\mathrm{K}^2 \sigma_e^2/2)^{1/2}$ is the radial component of
 the dust velocity dispersion, and $C_\mathrm{T}$ is a nondimensional
 constant.
The values of $C_\mathrm{T}$ are $C_\mathrm{T} = \pi$ for an ideal gas and
 $C_\mathrm{T} = 3.36$ for collisionless particles \citep{Toomre1964}; 
we adopted $C_\mathrm{T} = 3.36$.
In the most part of this paper, we adopt the condition for the GI as $Q<Q_\mathrm{cr}\simeq 2$ \citep{Michikoshi2016b}.

For comparison, we also examine the Roche criterion.
For the uniform dust layer perfectly coupled with the incompressible
gas, the Roche density is \citep{Sekiya1983} 
\begin{equation}
  \rho_\mathrm{R} \simeq 0.6 \frac{M_*}{a^3}.
  \label{eq:roche}
\end{equation}
We adopt this as the Roche density. Strictly speaking, the
coefficient of the Roche density depends on the vertical structure and
the equation of state of the dust layer. For instance, the coefficient
for the Gaussian dust density distribution is $0.78$ \citep{Yamoto2004}.
Furthermore in this paper we do not consider the incompressible gas.
Thus, Equation (\ref{eq:roche}) is considered as an order-of-magnitude estimate of
the Roche density for loosely coupling dust. From Equation (\ref{eq:roche}) we
define the nondimensional value $Q_\mathrm{R}$ \citep[e.g.,][]{Youdin2011} 
\begin{equation}
  Q_\mathrm{R} = \frac{h_\mathrm{d}}{h_\mathrm{cr}},
  \label{eq:roche_crit}
\end{equation}
where $h_\mathrm{d} = a \sigma_i$ is the dust scale height and
$h_\mathrm{cr} = \Sigma_\mathrm{d} / (\sqrt{\pi} \rho_\mathrm{R})$ is the critical scale height. If $Q_\mathrm{R} \lesssim 1$, the
dust layer tends to be unstable in the sense of the Roche criterion.

\section{Condition for Gravitational Instability of Dust Aggregates \label{sec:params}}

From $m_\mathrm{d}$ and $\rho_\mathrm{int}$, we can calculate the
 equilibrium value of $\sigma_e$ and $Q$ and draw the GI region on the
 $m_\mathrm{d}$--$\rho_\mathrm{int}$ plane where $Q < Q_\mathrm{cr}$.
We will show that a moderate mass is favorable for the GI. 
The existence and shape of the GI region depend on the various disk
 parameters.
In this section, we examine a general condition for the existence of
 a GI region in the $m_\mathrm{d}$--$\rho_\mathrm{int}$ plane, independent of the evolution of dust aggregates.
Whether an actual dust disk becomes gravitationally unstable
 depends on the mass-radius relation of the dust
 aggregates.
This issue will be discussed in Section \ref{sec:onset}, where we consider the evolution of the
 dust. 

 \subsection{Gravitational Instability Region in the Mass-Density Plane \label{sec:region}}

We begin by considering the equilibrium state.
We numerically calculate the equilibrium values of $\sigma_e$ and $\sigma_i$ by
 setting the right-hand sides of Equations (\ref{eq:evoe}) and
 (\ref{eq:evoi}) equal to zero.
Then, from Equation (\ref{eq:qvalue}), we calculate $Q$.

\begin{figure}
\begin{minipage}[b]{0.5\linewidth}
\begin{center}
(a) $\sigma_e$
\includegraphics[width=\textwidth]{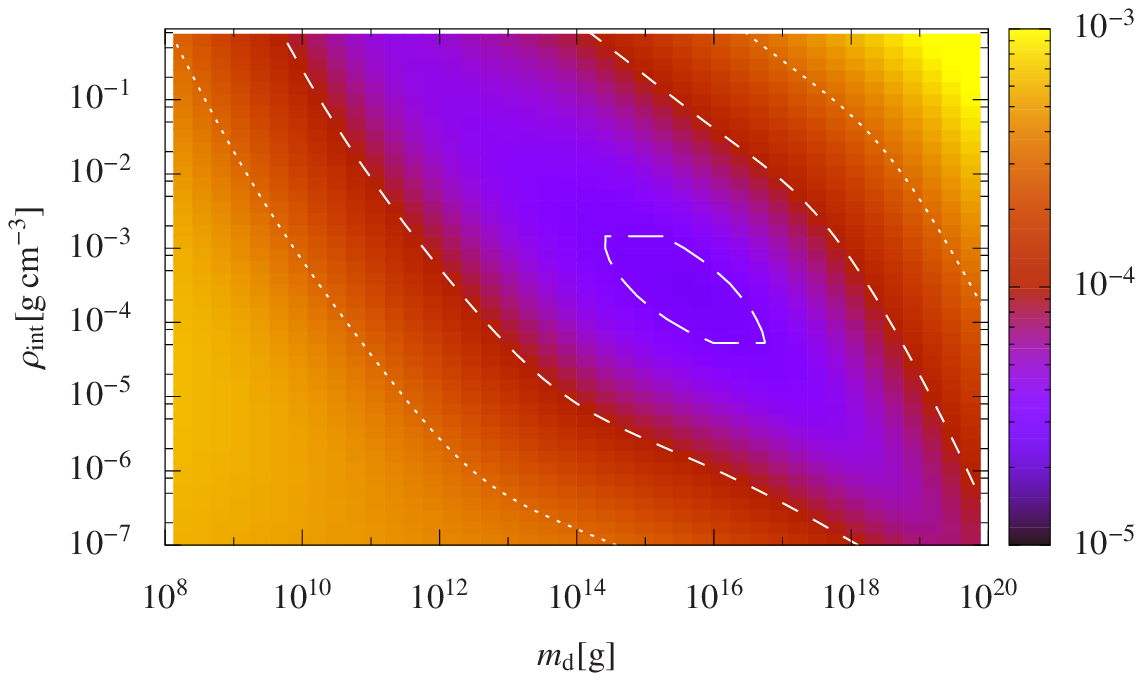}
\end{center}
\end{minipage} 
\begin{minipage}[b]{0.5\linewidth}
\begin{center}
(b) $\sigma_i$
\includegraphics[width=\textwidth]{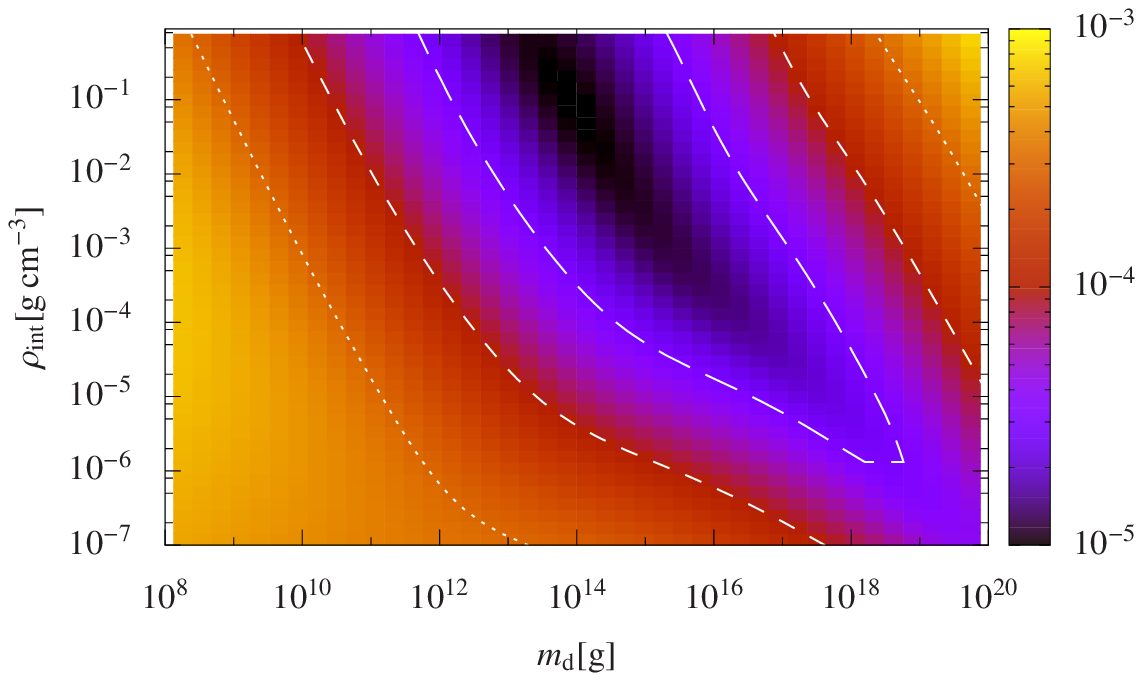}
\end{center}
\end{minipage} 
\caption{The equilibrium values of (a) eccentricity and (b) inclination on the $m_\mathrm{d}$--$\rho_\mathrm{int}$ plane for the fiducial model. The dashed, short-dashed, and dotted curves correspond to $3\times 10^{-5}$, $10^{-4}$, and $3 \times 10^{-4}$, respectively.
\label{fig:map_e_and_i}
} 
\end{figure}

Figure \ref{fig:map_e_and_i} shows the equilibrium $\sigma_e$ and
$\sigma_i$ for the fiducial model with $a = 5\, \mathrm{AU}$, $\alpha =
10^{-3}$, and $f_\mathrm{g} = 1$. In this model, $\sigma_e$ and $\sigma_i$
range from $10^{-5}$ to $10^{-3}$. Basically $\sigma_e$ and $\sigma_i$
have the similar dependencies on $m_\mathrm{d}$ and
$\rho_\mathrm{int}$. In most of the parameters regime, $\sigma_i$
smaller than $\sigma_e$. The ratio of $\sigma_i$ to $\sigma_e$ is
discussed in detail below. Around the region where $m_\mathrm{d} \sim
10^{15}\,\mathrm{g}$ and $\rho_\mathrm{int}\sim 10^{-4} \,
\mathrm{g}\,\mathrm{cm}^{-3}$, $\sigma_e$ has the smallest value of
about $10^{-5}$. The GI is likely to occur around this region in the
fiducial model.

\begin{figure}
\plotone{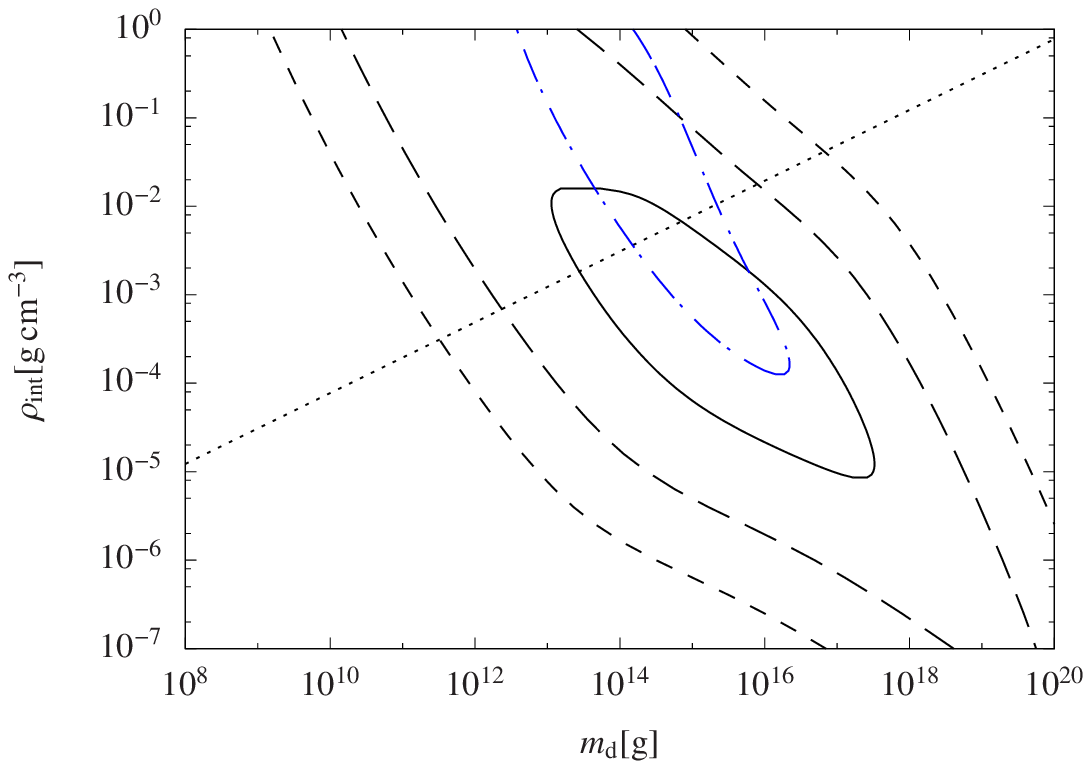}
\caption{
Gravitational instability region on the
 $m_\mathrm{d}$--$\rho_\mathrm{int}$ plane for the fiducial model.
The solid, dashed, and short dashed curves correspond to
 $Q = 1$, $2$, and $4$, respectively.  
The dotted line denotes the evolution track of the self-gravity
 compression.
 The dashed-dotted curve corresponds to $Q_\mathrm{R} = 2$.
\label{fig:q_contour}
} 
\end{figure}

We investigate $Q$ on the $m_\mathrm{d}$--$\rho_\mathrm{int}$ plane.
Figure \ref{fig:q_contour} shows the result for the fiducial model.
The minimum value of $Q$ is at $m_\mathrm{d} = 6.3 \times 10^{15} \, \mathrm{g}$ and
 $\rho_\mathrm{int} = 1.4 \times 10^{-4} \, \mathrm{g}\,\mathrm{cm}^{-3}$, where
$Q_\mathrm{min} \simeq 0.6$,
 which is sufficiently small to allow the GI.
Note that $Q < Q_\mathrm{cr}$ when
 $\rho_\mathrm{int} = 1\times 10^{-7} \, \mathrm{g}\,\mathrm{cm}^{-3}$
 to $1 \,\mathrm{g}\,\mathrm{cm}^{-3}$.
Therefore, when the density is in the realistic range, the GI condition is
 inevitably satisfied during the evolution of the dust.

\begin{figure}
\plottwo{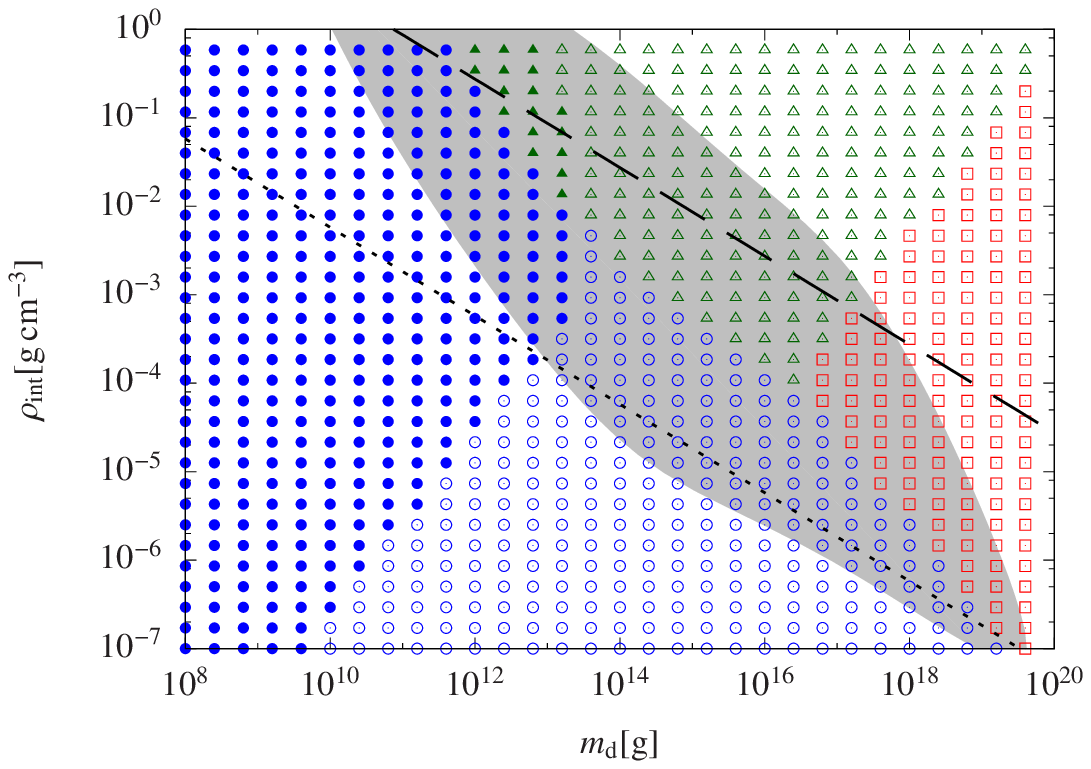}{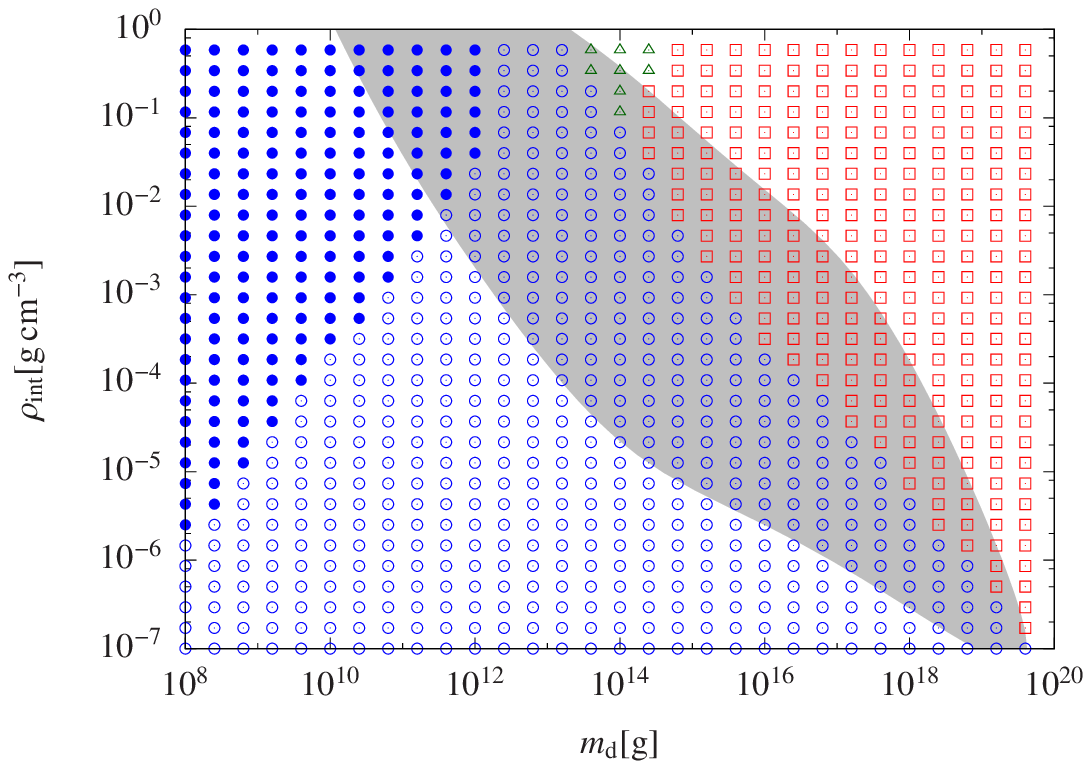}
\caption{
Dominant heating and cooling mechanisms for $\sigma_e$ (left) and
 $\sigma_i$ (right) for the fiducial model. 
The filled and open symbols represent gas drag and collisional damping
 for the dominant cooling process, respectively.  
The squares, circles, and triangles represent gravitational scattering,
 turbulent stirring, and turbulent scattering, respectively. 
The shaded region denotes the GI region where $Q < Q_\mathrm{cr}$. 
The dashed and dotted lines represent the approximated instability
 condition described by Equations (\ref{eq:mlow}) and (\ref{eq:mhigh}),
 respectively. 
}
\label{fig:source}
\end{figure}

The main heating and cooling mechanisms of $\sigma_e$ and $\sigma_i$ are
 shown on the $m_\mathrm{d}$--$\rho_\mathrm{int}$ plane for the fiducial
 model in Figure \ref{fig:source}.
In the region of low mass and low density, turbulent stirring is the
 dominant heating mechanism because the gas drag is strong. 
In the region of high mass and high density, turbulent scattering is the
 dominant heating mechanism, and in the region of high mass and low density,
 gravitational scattering is the dominant heating mechanism.
The heating rate due to gravitational scattering is proportional
to the scattering cross-section, which is about
$(G m_\mathrm{d}/v_\mathrm{ran}^2)^2 $ \citep[e.g.,][]{Ida1990}.
In the high mass region, the random velocity is approximately given by the
escape velocity because collisional damping and gravitational
scattering are dominant. Thus, Figure \ref{fig:map_e_and_i} shows that
in this region, the random velocity decreases with decreasing
$\rho_\mathrm{int}$. Therefore, gravitational scattering is stronger
for the lower density if we fix the mass.
The region of turbulent scattering for $\sigma_i$ is smaller than that for
 $\sigma_e$ because the scattering rate for $\sigma_i$ is smaller than it is for
 $\sigma_e$, as shown in Equation (\ref{eq:heat_i_gravsca}). 

We compare the Roche criterion with the Toomre criterion.
In the fiducial model, there is no region for $Q_\mathrm{R}< 1$. As shown in
Figure \ref{fig:q_contour}, there is the region for $Q_\mathrm{R} < 2$, but it is smaller than that
for $Q < 2$. Thus, the Roche criterion is harder to be satisfied than
the Toomre criterion. The region for $Q_\mathrm{R} < 2$ relatively extends
towards high $\rho_\mathrm{int}$. In this region, the main heating
source is turbulent scattering. The heating rate of the inclination
due to turbulent scattering is small, which leads to a thin dust
layer. Thus, there the dust layer is more likely to be unstable in the
sense of the Roche criterion. In the following we use the Toomre
criterion, which is more optimistic.

\subsection{Dynamical Properties of Dust Aggregates \label{sec:dynamics_prop}}

\begin{figure}
\begin{minipage}[b]{0.5\linewidth}
\begin{center}
(a) $v_\mathrm{ran}/v_\mathrm{esc}$
\includegraphics[width=\textwidth]{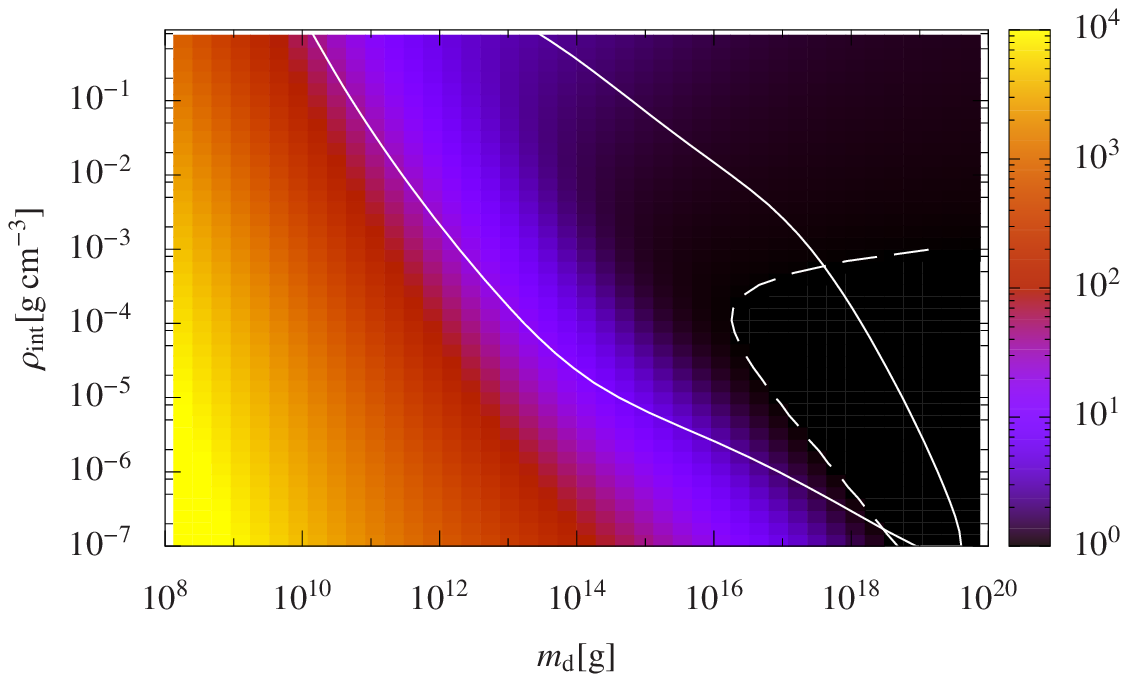}
\end{center}
\end{minipage} 
\begin{minipage}[b]{0.5\linewidth}
\begin{center}
(b) $v_\mathrm{ran}/(\eta v_\mathrm{K})$
\includegraphics[width=\textwidth]{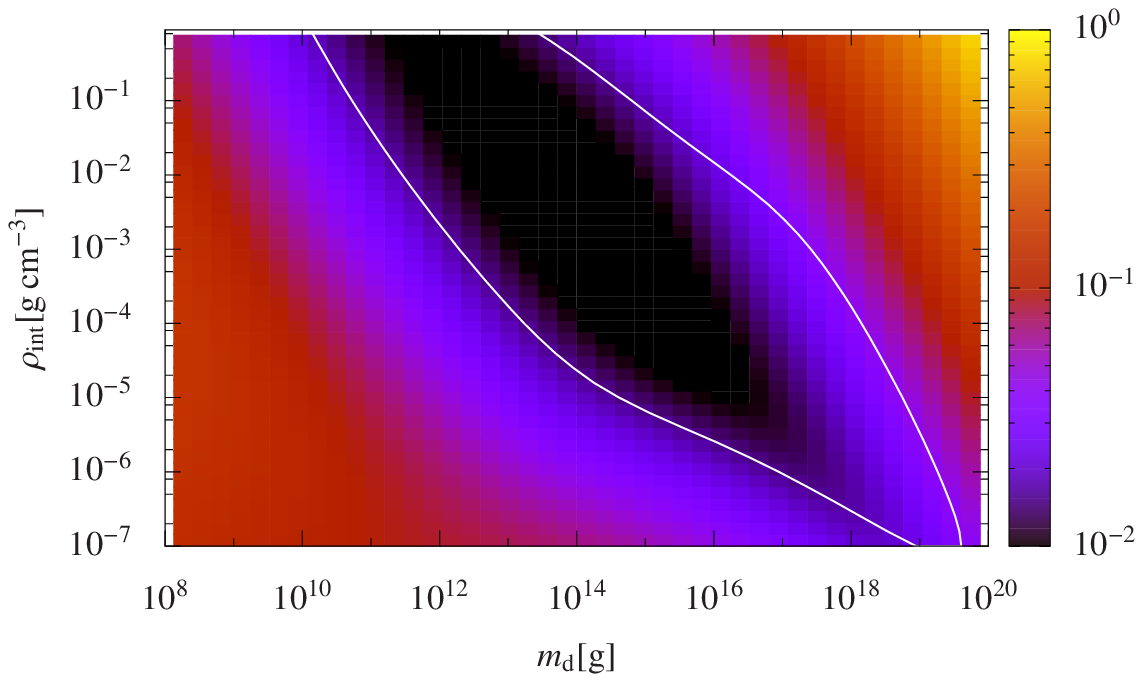}
\end{center}
\end{minipage} 
\begin{minipage}[b]{0.5\linewidth}
\begin{center}
(c) $v_\mathrm{ran}/v_\mathrm{H}$
\includegraphics[width=\textwidth]{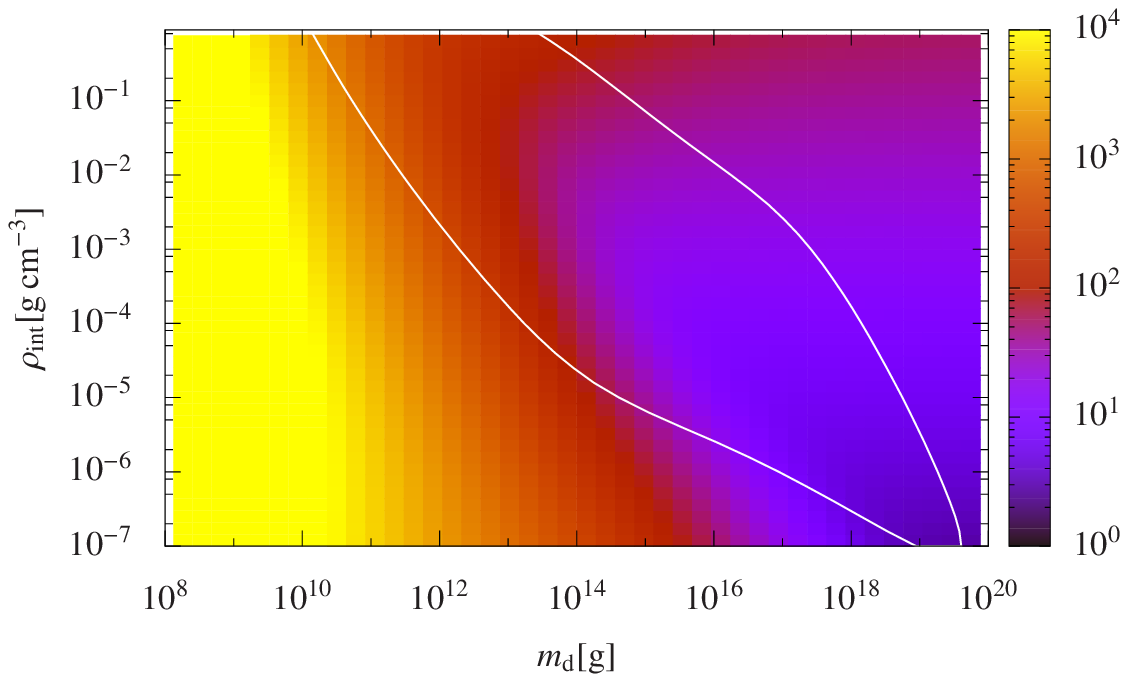}
\end{center}
\end{minipage} 
\begin{minipage}[b]{0.5\linewidth}
\begin{center}
(d) $\sigma_i/\sigma_e$
\includegraphics[width=\textwidth]{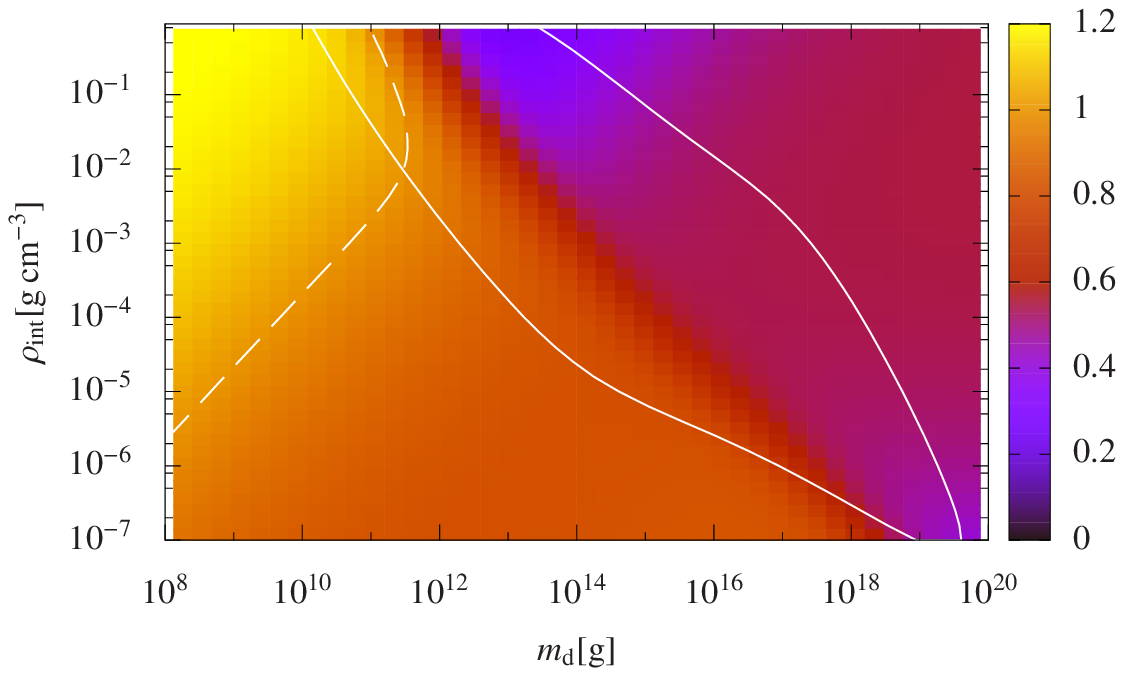}
\end{center}
\end{minipage} 
\begin{minipage}[b]{0.5\linewidth}
\begin{center}
(e) $C_\mathrm{D}$
\includegraphics[width=\textwidth]{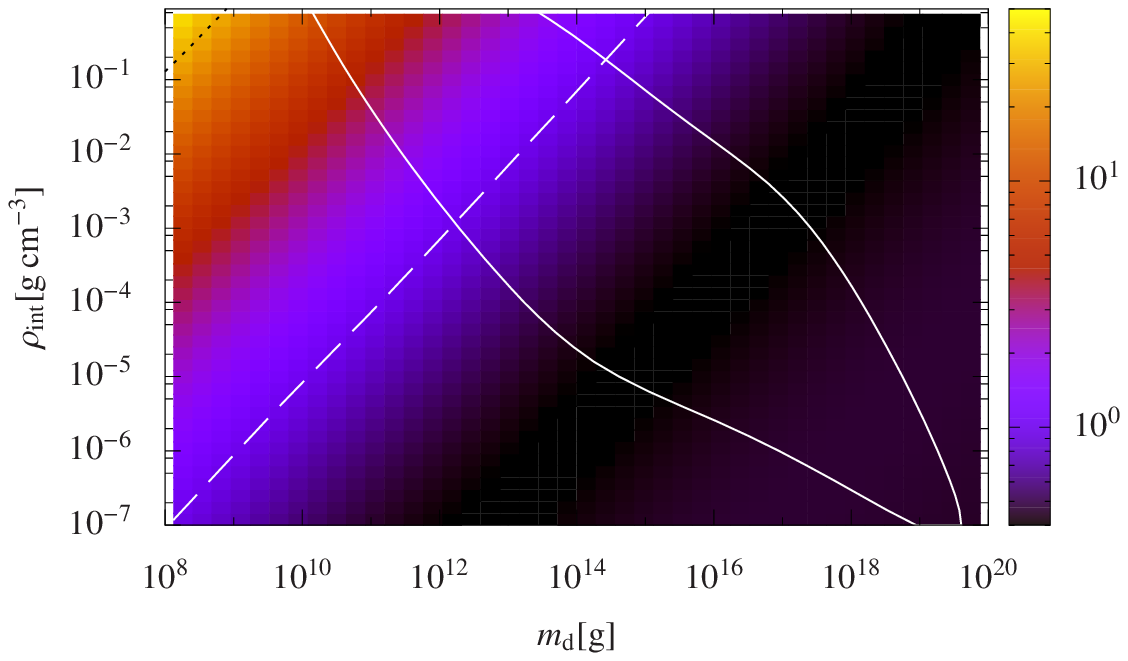}
\end{center}
\end{minipage} 
\begin{minipage}[b]{0.5\linewidth}
\begin{center}
(f) $S$
\includegraphics[width=\textwidth]{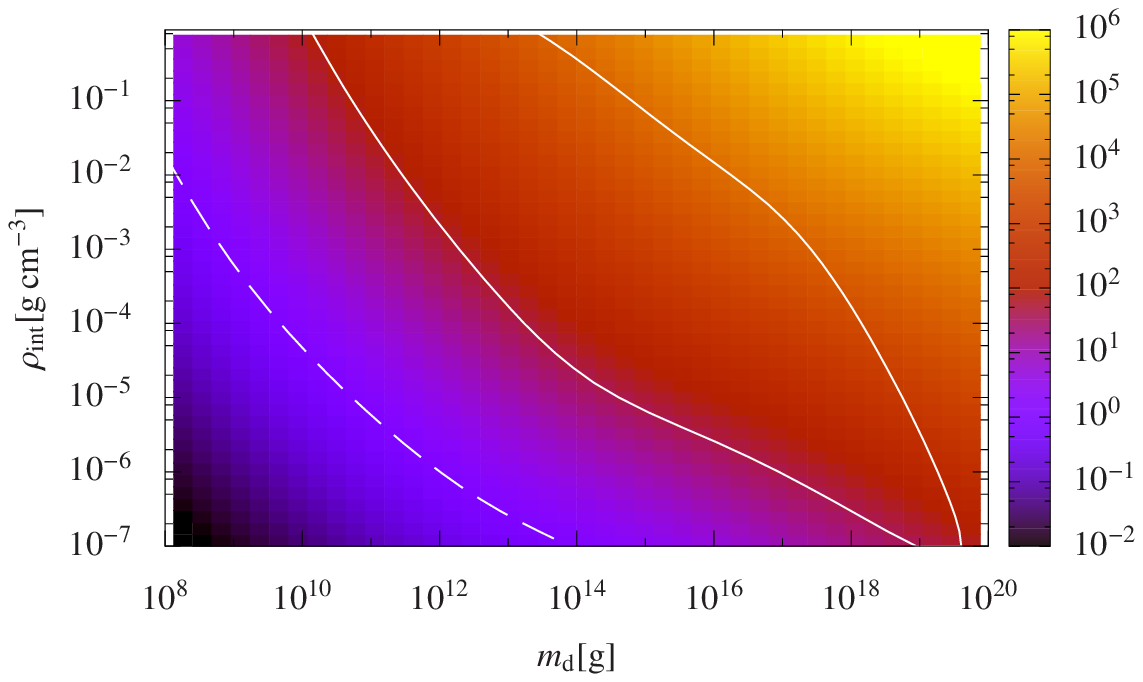}
\end{center}
\end{minipage} 
\caption{
Dynamical parameters of dust aggregates
 (a) $v_\mathrm{ran}/v_\mathrm{esc}$,
 (b) $v_\mathrm{ran}/\eta v_\mathrm{K}$,
 (c) $v_\mathrm{ran}/ v_\mathrm{H}$,
 (d) $\sigma_i/\sigma_e$,
 (e) $C_\mathrm{D}$, and
 (f) $S$ on the $m_\mathrm{d}$--$\rho_\mathrm{int}$ plane for the
 fiducial model. 
The solid and dashed curves show $Q = Q_\mathrm{cr}$ and unity, respectively. 
The dotted line in the panel (e) show $\mathrm{Re}=1$ ($C_\mathrm{D}=27.6$).
}
\label{fig:various_ratio}
\end{figure}

We examine the dust aggregate dynamics in detail to clarify the physical
 background of the GI. 
Figure \ref{fig:various_ratio}a shows the ratio of the random velocity
 to the surface escape velocity, $v_\mathrm{esc}=\sqrt{2 G m_\mathrm{d}/r_\mathrm{c}}$.
If this ratio is less than unity, gravitational focusing is effective.
In the low-mass region, the ratio is sufficiently larger than unity, which
 means gravitational focusing is negligible.
In this case, the collisional cross-section is well approximated by the geometrical cross-section, and the condition for runaway growth is not satisfied \citep[e.g.,][]{Ohtsuki1993, Kokubo1996}.
In the high-mass region, the escape velocity is comparable to the random velocity, 
and thus, the collisional cross-section is enhanced by a factor of about $2$.

When calculating $u$ (Equation (\ref{eq:relative})), if $v_\mathrm{ran}/(\eta v_\mathrm{K})$ is small, then $v_\mathrm{ran}$ is negligible.
Figure \ref{fig:various_ratio}b shows $v_\mathrm{ran}/(\eta v_\mathrm{K})$.
Other than when the mass and density are both high,
 this ratio is less than unity, and we can safely adopt the
 approximation $u \simeq \eta v_\mathrm{K}$. 

 The ratio of the random velocity to the Hill velocity, $v_\mathrm{H} = r_\mathrm{H}\Omega$, determines the regime of gravitational scattering, that is, the shear-dominated or dispersion-dominated regime \citep{Weidenschilling1989}.
Figure \ref{fig:various_ratio}c shows $v_\mathrm{ran}/v_\mathrm{H}$.
In the entire region, $v_\mathrm{ran}$ is larger than $v_\mathrm{H}$, which
 indicates that the random velocity is dispersion-dominated.

Figure \ref{fig:various_ratio}d shows the ratio $\sigma_i/\sigma_e$,
 which is determined by the heating and cooling mechanisms.
Figure \ref{fig:source} shows the main heating and cooling mechanisms on the
 $m_\mathrm{d}$--$\rho_\mathrm{int}$ plane. 
For the region where the main heating and cooling mechanism is
 gravitational scattering and collisions, respectively,
 $\sigma_i/\sigma_e$ is about $0.4\mbox{--}0.6$,  
 because gravitational scattering results in $\sigma_i/\sigma_e \simeq 0.5$
 \citep{Ida1992}.
In the region with turbulent drag and gas drag, $\sigma_i/\sigma_e$ is
 larger than unity, and the damping rate of $\sigma_e$ due to gas drag
 is larger than that of $\sigma_i$.
Thus, $\sigma_i$ is larger than $\sigma_e$.
The ratio $\sigma_i/\sigma_e$ is about $0.8$ where turbulent drag and 
 collisions are dominant.
In the region where turbulent scattering and collisions prevail,
 $\sigma_i/\sigma_e$ is less than $0.2$, and the heating rate of
 $\sigma_i$ due to turbulent scattering is smaller 
 than that of $\sigma_e$. 
This leads to a smaller value for $\sigma_i/\sigma_e$ \citep{Yang2012, Kobayashi2016}.

Figure \ref{fig:various_ratio}e shows the drag coefficient
 $C_\mathrm{D}$.
There is no Epstein regime in Figure \ref{fig:various_ratio}e. The Epstein
regime appears when we consider the small dust aggregates with $m_\mathrm{d} <
10^7 \mathrm{g}$. For $\mathrm{Re} < 1$ ($C_\mathrm{D} > 27.6$), the
Stokes law is a good approximation.
In the region where the mass is small and the density is large, the gas
drag obeys Stokes' law.
On the other hand, for massive dust aggregates, the gas drag
obeys Newton's law.
Thus, in this region, $C_\mathrm{D}$ is almost constant, at
 about $0.4 \mbox{--} 0.5$. 

The Stokes number is shown in Figure \ref{fig:various_ratio}f.
For the low-mass, low-density region, $S$ is less than unity, which means the
 dust is well coupled with the gas.
In the GI region, $S$ is sufficiently larger than unity the coupling does not occur.

We now consider the main heating and cooling processes shown in Figure \ref{fig:source}.
In the low-mass, high-density region, gas drag is the main cooling mechanism, while in the high-mass, low-density region, collisions dominate. 
Gas drag obeys Stokes' law in the low-mass, high-density region,
 as shown in Figure \ref{fig:various_ratio}e. 
In this case, the damping rate due to gas drag is
 $\propto C_\mathrm{D} r_\mathrm{c}^2/m_\mathrm{d} \propto r_\mathrm{c}
 / m_\mathrm{d} \propto m_\mathrm{d}^{-2/3} \rho_\mathrm{int}^{-1/3}$. 
Similarly, neglecting gravitational focusing, the damping rate due
 to collisions is
 $\propto r_\mathrm{c}^2/ m_\mathrm{d} \propto m_\mathrm{d}^{-1/3}
 \rho_\mathrm{int}^{-2/3}$.  
Thus, for larger $m_\mathrm{d}$ and smaller $\rho_\mathrm{int}$, damping by collisions is more important than that by gas drag.

The region where gas drag is dominant for $\sigma_e$ is larger than that
 for $\sigma_i$. 
As shown in Figure \ref{fig:various_ratio}b, in most areas,
 $\sigma_e,\sigma_i < \eta$. 
In this case, the damping rate due to gas drag can be approximated as
 $ (\mathrm{d} \sigma_e^2/\mathrm{d}t)_\mathrm{drag} \simeq -
 (3/t_\mathrm{s}) \sigma_e^2 $ and
 $ (\mathrm{d} \sigma_i^2/\mathrm{d}t)_\mathrm{drag} \simeq -
 (1/t_\mathrm{s}) \sigma_i^2$. 
The damping time for $\sigma_e$ is $t_\mathrm{s}/3$, which is shorter
 than that for $\sigma_i$. 
On the other hand, the damping times due to collisions for
 $\sigma_e$ and $\sigma_i$ are the same. 
Thus, the gas drag region for $\sigma_e$ is larger than that for $\sigma_i$.

\subsection{Critical Turbulent Strength for Gravitational Instability \label{sec:cond}} 

By a similar way to that used in Paper I, we derive the critical $\alpha$ for
the existence of the GI region. First we focus on the lower mass
boundary of the GI region around $\rho_\mathrm{int} \simeq 10^{-6}\,\mathrm{g}\,\mathrm{cm}^{-3}$.
In the low-density region, the main heating and cooling mechanisms are
turbulent stirring and collisions, respectively.
In this case, the equilibrium $\sigma_e$ is approximated by
$(\mathrm{d}\sigma_e^2/\mathrm{d}t)_\mathrm{turb,stir} +
(\mathrm{d}\sigma_e^2/\mathrm{d}t)_\mathrm{col} \simeq 0$.
We neglect the second term in Equation (\ref{eq:pcoleq}) since
$v_\mathrm{ran} > v_\mathrm{esc}$ and gravitational focusing is not
effective as shown in Figure \ref{fig:various_ratio}a.
Furthermore, as shown in Figure \ref{fig:various_ratio}d,
$\sigma_i/\sigma_e$ is almost constant in the low-density region.
Thus, assuming $\sigma_i/\sigma_e = 0.71$, we evaluate the first term of
Equation (\ref{eq:pcoleq}) and obtain $P_\mathrm{col} \simeq 2.13
\tilde r_\mathrm{c}^2$. Substituting this into Equation
(\ref{eq:damp_col}), we obtain the approximated $(\mathrm{d}\sigma_e^2/\mathrm{d}t)_\mathrm{col}$.
Figure \ref{fig:various_ratio}f shows $S \gg 1$, where
we obtain $(\mathrm{d} \sigma_e^2/\mathrm{d}t)_\mathrm{turb,stir}
\simeq 4 \tau_\mathrm{e} v_\mathrm{t}^2 \Omega / 3 v_\mathrm{K}^2 S^2$
from Equation (\ref{eq:turbheat_e}).
We find that $v_\mathrm{ran} < \eta v_\mathrm{K}$ in Figure
\ref{fig:various_ratio}b.
Thus, from Equation (\ref{eq:relative}), we approximate $u \simeq \eta
v_\mathrm{K}$. Substituting this into Equation (\ref{eq:ts1}), we
calculate $S$ and finally obtain the approximated
$(\mathrm{d}\sigma_e^2/\mathrm{d}t)_\mathrm{turb,stir}.$
From Figure \ref{fig:various_ratio}e, the Newton's drag is a good
approximation, thus we assume $C_\mathrm{D} \simeq 0.5$.
Based upon these approximations, we calculate $\sigma_e$ and obtain 
\begin{equation}
Q \simeq 3.24 \times 10^{-2} \frac{\alpha^{1/2} \eta \tau_\mathrm{e}^{1/2}
C_\mathrm{D} M_* \Sigma_\mathrm{g} }{C_\mathrm{col}^{1/2} a^2
m_\mathrm{d}^{1/6} \rho_\mathrm{int}^{1/3} \Sigma_\mathrm{d}^{3/2}}. 
\end{equation}
From $Q < Q_\mathrm{cr}$, we obtain the inequality
\begin{equation}
m_\mathrm{d} \gtrsim
m_\mathrm{low} =
1.16\times 10^{-9}
 \frac{\alpha^3  \eta^6\tau_\mathrm{e}^3 C_\mathrm{D}^6 M_*^6 \Sigma_\mathrm{g}^6 }{
   C_\mathrm{col}^3 Q_\mathrm{cr}^6 a^{12} \rho_\mathrm{int}^2 \Sigma_\mathrm{d}^9 }. 
\label{eq:mlow}
\end{equation}
Next we focus on the upper mass boundary of the GI region around $\rho_\mathrm{int} \simeq 10^{-1}\,\mathrm{g}\,\mathrm{cm}^{-3}$.
In the high-mass, high-density region,
the main heating and cooling mechanisms are turbulent scattering and collisions, respectively. 
We obtain the equilibrium value for $\sigma_e$ from
 $(\mathrm{d}\sigma_e^2/\mathrm{d}t)_\mathrm{turb,grav} +
 (\mathrm{d}\sigma_e^2/\mathrm{d}t)_\mathrm{col} = 0$. 
In evaluating $(\mathrm{d}\sigma_e^2/\mathrm{d}t)_\mathrm{col}$, we adopt the same approximations described above.
From $Q < Q_\mathrm{cr}$, we obtain the upper limit of $m_\mathrm{d}$:
\begin{equation}
m_\mathrm{d} \lesssim
m_\mathrm{high} =
4.04 \times 10^5 \frac{C_\mathrm{col}^3 Q_\mathrm{cr}^6
 \Sigma_\mathrm{d}^9}{\alpha^3 C_\mathrm{turb}^3 \rho_\mathrm{int}^2
 \Sigma_\mathrm{g}^6}. 
\label{eq:mhigh}
\end{equation}
In Figure \ref{fig:source}, $m_\mathrm{low}$ and $m_\mathrm{high}$ are
 plotted, and they are in rough agreement with the numerical results. 

For the dust aggregates to trigger the GI, it is necessary that $m_\mathrm{low} < m_\mathrm{high}$. 
If this inequality is not satisfied, the GI region does not exist. 
Thus, we derive the following condition for the GI:
\begin{equation}
\alpha < \alpha_\mathrm{cr,1} =
2.65\times10^2 \frac{ C_\mathrm{col} Q_\mathrm{cr}^2 a^2
  \Sigma_\mathrm{d}^3}{\eta \tau_\mathrm{e}^{1/2} C_\mathrm{turb}^{1/2} C_\mathrm{D}
 M_* \Sigma_\mathrm{g}^2 }. 
\end{equation}
Using the disk model, we rewrite $\alpha_\mathrm{cr,1}$ as
\begin{eqnarray}
\alpha_\mathrm{cr,1} & = & 8.75 \times 10^{-3} \tau_\mathrm{e}^{-1/2}
 f_\mathrm{g} \left(\frac{\gamma}{0.018}\right)^3 \left(\frac{a}{1\, \mathrm{AU}} \right)^{\beta_\mathrm{t}-\beta_\mathrm{g}+1}
 \left(\frac{T_1}{120}\right)^{-1} \left(\frac{M_*}{M_\odot}\right)^{-1}
 \nonumber \\ 
& & \times \left(\frac{2.39 \beta_\mathrm{g} + 1.20 \beta_\mathrm{t} + 3.59}{7.70} \right)^{-1} 
  \left(\frac{C_\mathrm{turb}}{3.1 \times 10^{-2}} \right) ^{-1/2}.  
\label{eq:cond}
\end{eqnarray}
As shown in Section \ref{sec:onsetgi}, this condition agrees well with
 the numerical results.

In the fiducial model ($\beta_\mathrm{g} = 3/2$ and $\beta_\mathrm{t} = 3/7$),
 we find a weak dependence on $a$: $\propto a^{-1/14}$.
Note that for the optically thin minimum-mass solar nebular model
 ($\beta_\mathrm{t}=1/2$ and $\beta_\mathrm{g} = 3/2$)
 \citep{Hayashi1981, Hayashi1985}, this dependence vanishes completely.
Thus, for realistic disk models, the dependence on $a$ is generally
 weak.

\section{Gravitational Instability with Evolution of Dust \label{sec:onset}}
 
In the previous section, we investigated a condition on the dust aggregate that would bring about the GI; in this section, we examine whether the GI occurs as dust aggregates evolve in a protoplanetary disk.

\subsection{Dependence on Disk Parameters \label{sec:onsetgi}}

\begin{figure}
  \begin{minipage}[b]{0.5\linewidth}
	\begin{center}
		(a) 
		\plotone{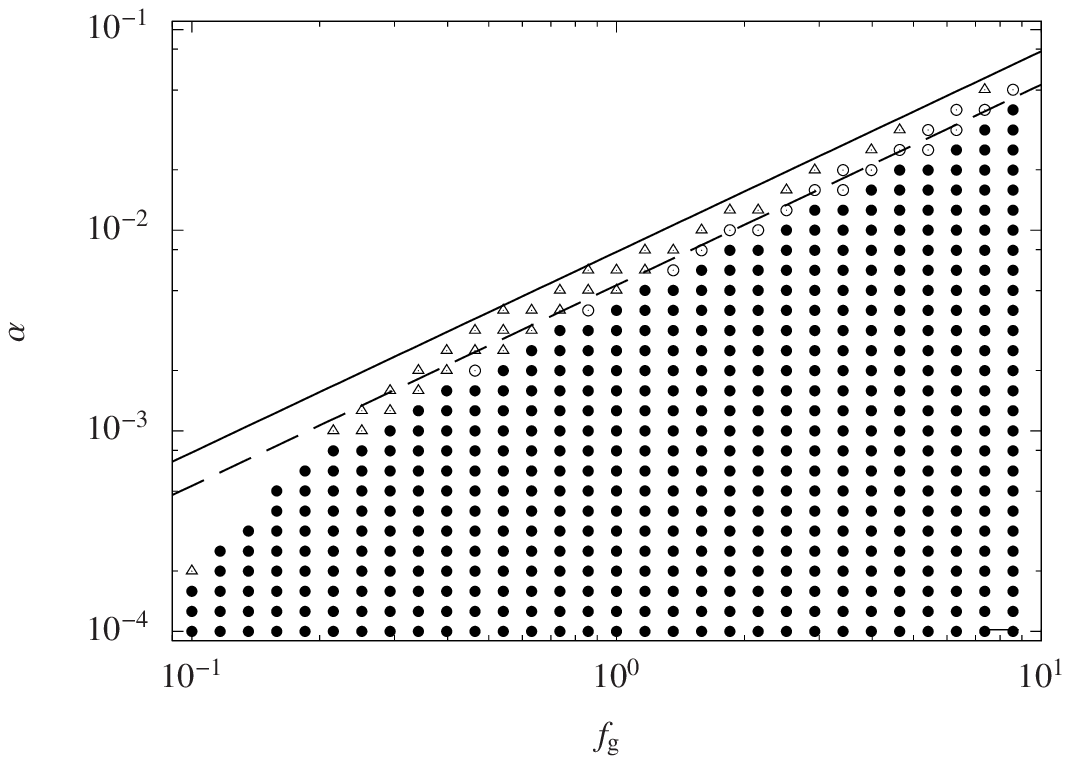}
	\end{center}
  \end{minipage} 
  \begin{minipage}[b]{0.5\linewidth}
	\begin{center}
	  (b) 
	  \plotone{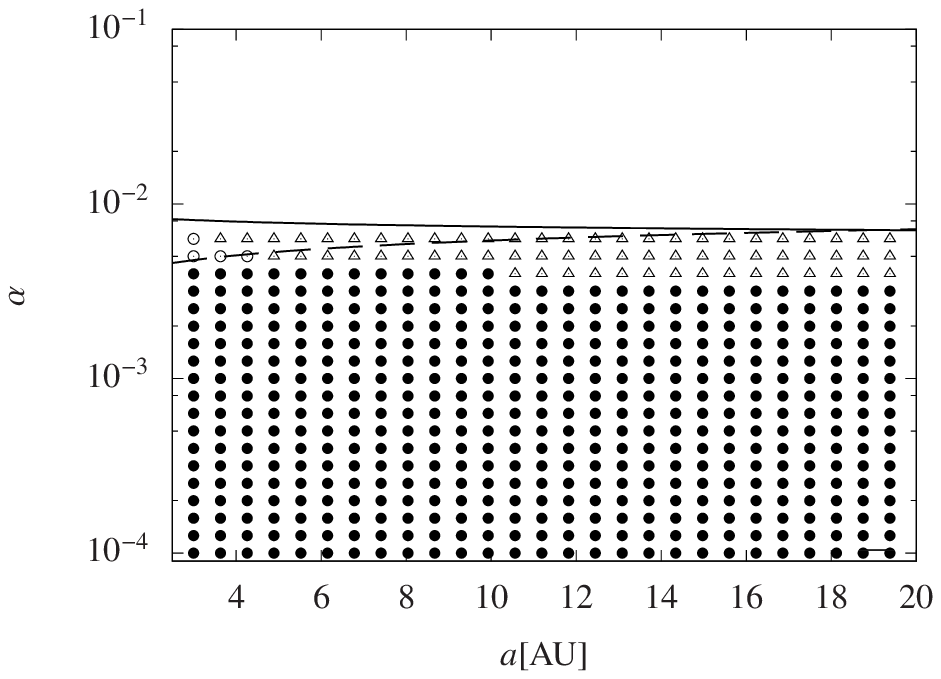}
	\end{center}
  \end{minipage} 
  \begin{minipage}[b]{0.5\linewidth}
	\begin{center}
		(c) 
		\plotone{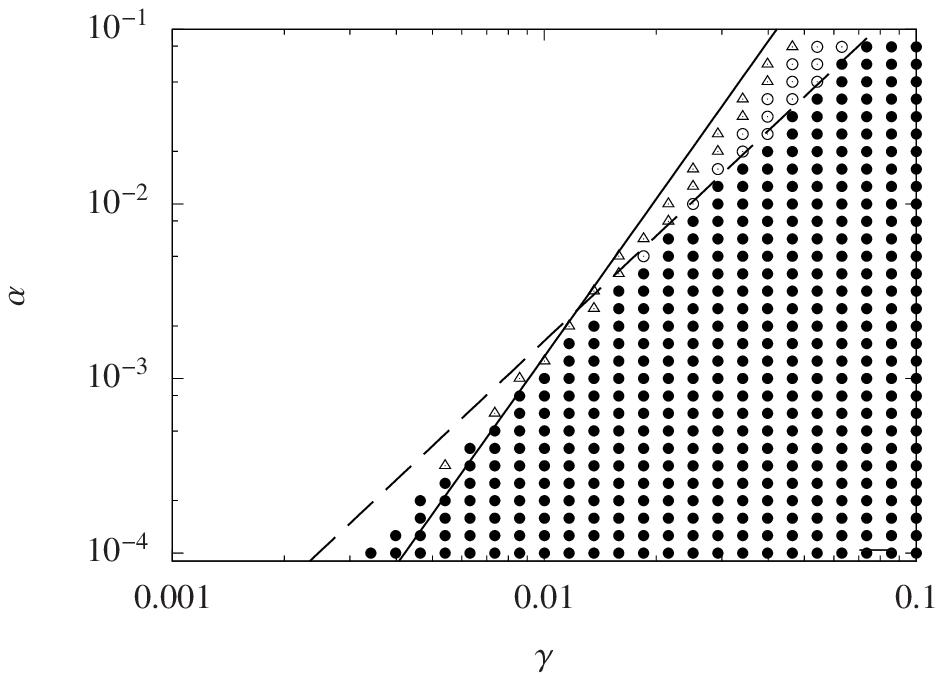}
	\end{center}
  \end{minipage} 
  \begin{minipage}[b]{0.5\linewidth}
	  \begin{center}
 		 (d) 
		  \plotone{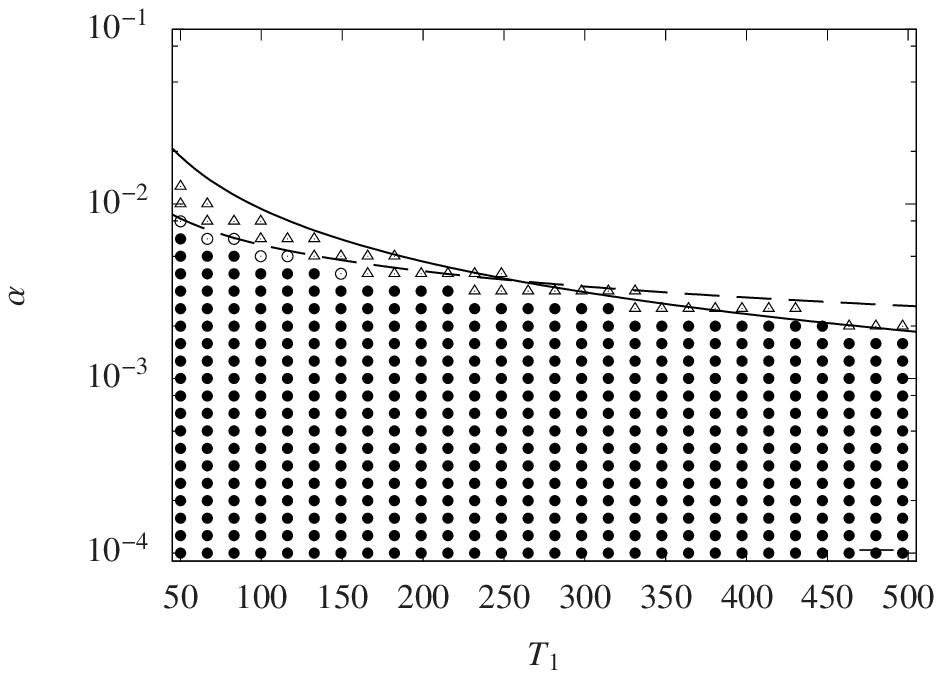}
	  \end{center}
  \end{minipage} 
  \begin{minipage}[b]{0.5\linewidth}
	  \begin{center}
		  (e) 
		  \plotone{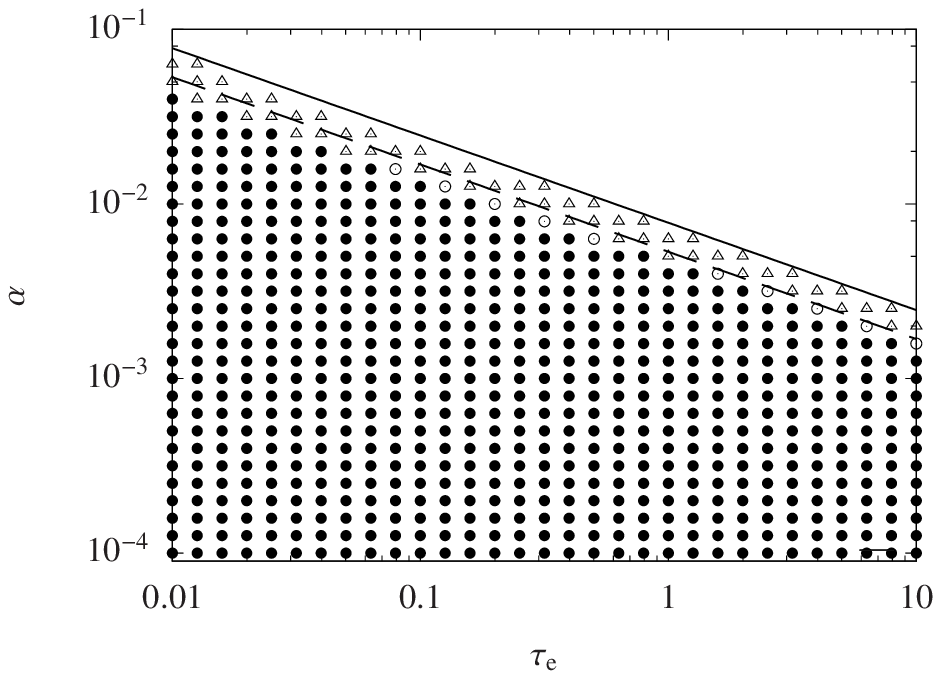}
	  \end{center}
  \end{minipage} 
  \begin{minipage}[b]{0.5\linewidth}
	  \begin{center}
		  (f) 
		  \plotone{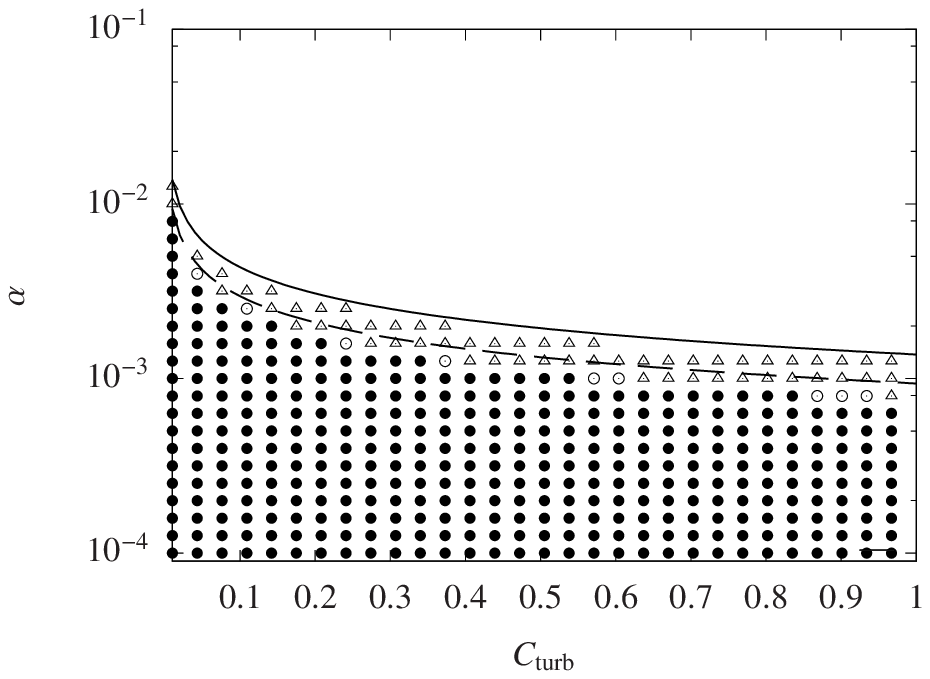}
	  \end{center}
  \end{minipage} 
  \caption{ Critical viscous parameter $\alpha$ for the GI versus (a) $f_\mathrm{g}$, (b) $a$, (c) $\gamma$, (d) $T_1$, (e) $\tau_\mathrm{e}$, and (f) $C_\mathrm{turb}$.
  All points show the existence of the GI region in the area $10^8 <m_\mathrm{d} < 10^{20} \mathrm{g}$ and $10^{-6} \, \mathrm{g}\, \mathrm{cm}^{-3} < \rho_\mathrm{int} < 1 \, \mathrm{g}\, \mathrm{cm}^{-3}$. 
  Open triangles indicate that the dust evolution does not cross the GI region. 
  Open circles indicate where the dust evolution crosses the GI region with the equilibrium random velocity but not with the nonequilibrium random velocity, while filled circles show where the dust evolution crosses the GI region with both equilibrium and nonequilibrium random velocities. 
  The solid and dashed curves show $\alpha_\mathrm{cr,1}$ (Equation (\ref{eq:cond})) and $\alpha_\mathrm{cr,2}$ (Equation (\ref{eq:cond2})). 
  \label{fig:critical_alpha}
  }
\end{figure}

First, assuming the equilibrium random velocity, we examine the dependence of the critical $\alpha$ for the GI on the disk parameters.
We adopt the dust evolution described in Section \ref{sec:evotrack} and check whether its track crosses the GI region.
Figure \ref{fig:critical_alpha}a shows the dependence on $f_\mathrm{g}$.
The GI is more likely with larger $f_\mathrm{g}$ and smaller $\alpha$. 
The preference for small $\alpha$ occurs because turbulence is the main heating mechanism. 
From Equation (\ref{eq:okuzumi_formula2013}), when the gas surface density 
 ($f_\mathrm{g}$) is large, the heating rate due to turbulent scattering is also large. 
In our model, the dust surface density also increases with
 $f_\mathrm{g}$, since the dust-to-gas density ratio $\gamma$ is fixed.
When the dust surface density is large, this leads to strong self-gravity and a high collision frequency.
Among these competing effects, the increase in self-gravity and collision frequency are dominant.
Thus, the GI is more likely to occur when $f_\mathrm{g}$ is large.

\begin{figure}
\plotone{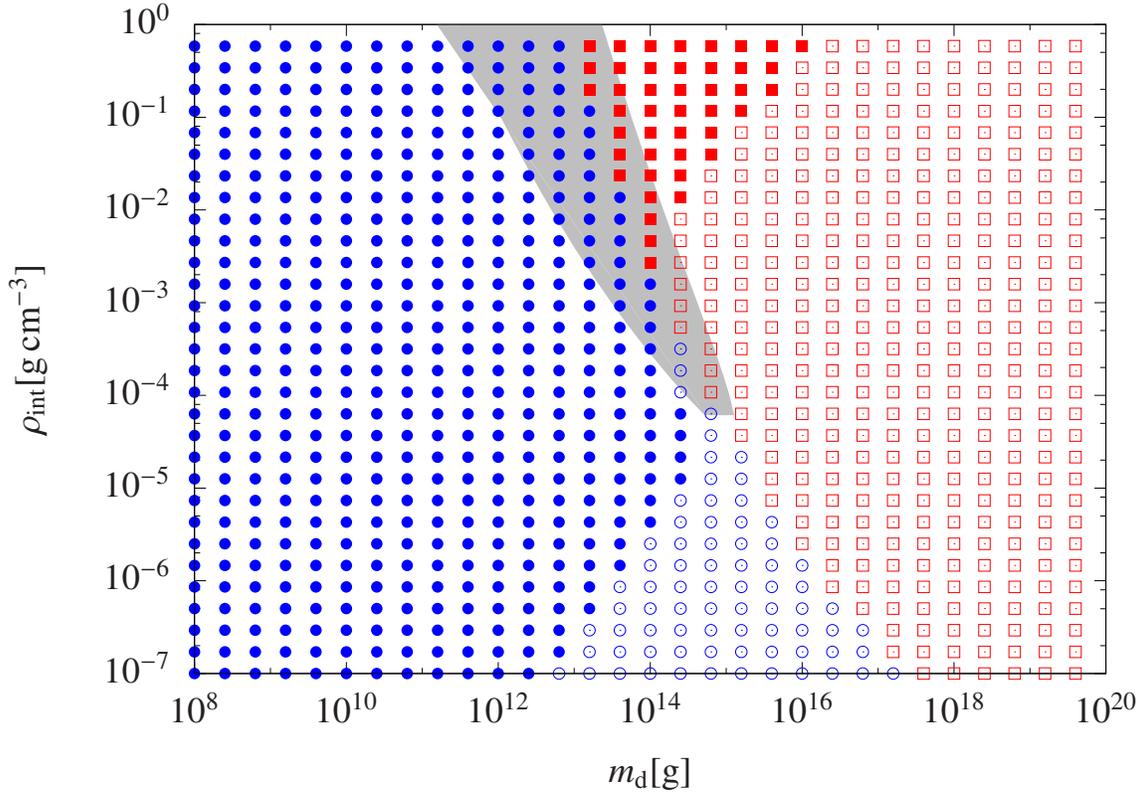}
\caption{
Main heating and cooling mechanisms for $\sigma_e$ with
 $\alpha = 1\times 10^{-4}$ and $f_\mathrm{g}=0.1$. 
The other parameters are the same as those of the fiducial model.
The symbols are the same as those in Figure \ref{fig:source}. 
The shaded region is the GI region.
}
\label{fig:source3}
\end{figure}

We found that the condition for the existence of the GI region (Equation
 (\ref{eq:cond})) agrees well with the numerical results for
 $f_\mathrm{g} > 0.3$, while it overestimates $\alpha$ for
 $f_\mathrm{g} < 0.3$.
To determine the reason for this, in Figure
 \ref{fig:source3}, we plotted the main heating and cooling mechanisms
 for $\alpha = 10^{-4}$ and $f_\mathrm{g} = 0.1$.  
For low $f_\mathrm{g}$, turbulent scattering is insignificant
 because its heating rate is proportional
 to $\Sigma_\mathrm{g}^2$. 
In deriving Equation (\ref{eq:cond}), we assumed that the upper
 boundary of the GI region is determined by a balance between 
 turbulent scattering and collisions. 
However, this assumption breaks down, and thus, in this parameter region, Equation (\ref{eq:cond}) differs from the numerical results. 

As expected from Equation (\ref{eq:cond}), the dependence on $a$ is weak.
From the numerical results shown in Figure \ref{fig:critical_alpha}b, the existence of the GI region is independent of $a$.
The value of $\alpha$ required to cross the GI region slightly decreases
 with $a$, but its dependence is very weak. 

As shown in Figure \ref{fig:critical_alpha}c, the dependence on the
 dust-to-gas ratio $\gamma$ is strong.
The GI easily takes place for larger values of $\gamma$.
The critical $\alpha$ for the existence of the GI
 region, $\alpha_\mathrm{cr,1}$ (Equation (\ref{eq:cond})),
 is proportional to $\gamma^3$, which agrees well with the numerical
 results. 
However, the dependence of the critical $\alpha$ on $\gamma$ for crossing the GI region is different; it is approximately proportional to $\gamma^2$. 
This dependence will be discussed below.
The difference from $\alpha_\mathrm{cr,1}$ increases with $\gamma$
 for  $\gamma \gtrsim 0.05$.  

The critical $\alpha$ decreases with increasing $T_1$, as shown in Figure
 \ref{fig:critical_alpha}d.
Increased temperatures lead to suppression of the GI. 
From Equation (\ref{eq:etamodel}), $\eta$ is proportional to $T_1$. 
The typical difference between the velocity of dust and that of gas is determined
 from $\eta v_\mathrm{K}$ because $v_\mathrm{ran} < \eta v_\mathrm{K}$. 
Thus, a higher $T_1$ means that the gas drag is strong, and this causes
 strong turbulent stirring and inhibits the GI. 
When $T_1 = 280$, the critical $\alpha$ that is often adopted is $0.4$
 times that when $T_1 = 120$.   

Figure \ref{fig:critical_alpha}e shows that the critical $\alpha$
 decreases with increasing $\tau_\mathrm{e}$ as $\tau_\mathrm{e}^{-1/2}$. 
A small $\tau_\mathrm{e}$ means that the duration of the fluctuations in the turbulent
 velocity are short. 
In this case, from Equation (\ref{eq:turbheat_e}), the heating rate due to
 turbulent stirring is small.
Thus, the GI tends to occur for smaller $\tau_\mathrm{e}$.

The critical $\alpha$ decreases with increasing $C_\mathrm{turb}$ as $C_\mathrm{turb}^{-1/2}$, as shown in Figure \ref{fig:critical_alpha}f.
From Equation (\ref{eq:okuzumi_formula2013}), the heating rate
 due to turbulent scattering decreases with increasing $C_\mathrm{turb}$. 
Thus, for smaller $C_\mathrm{turb}$, the critical $\alpha$ is larger.

\begin{figure}
\begin{minipage}[b]{0.5\linewidth}
\begin{center}
(a) $\beta_\mathrm{g}=0.5$
\plotone{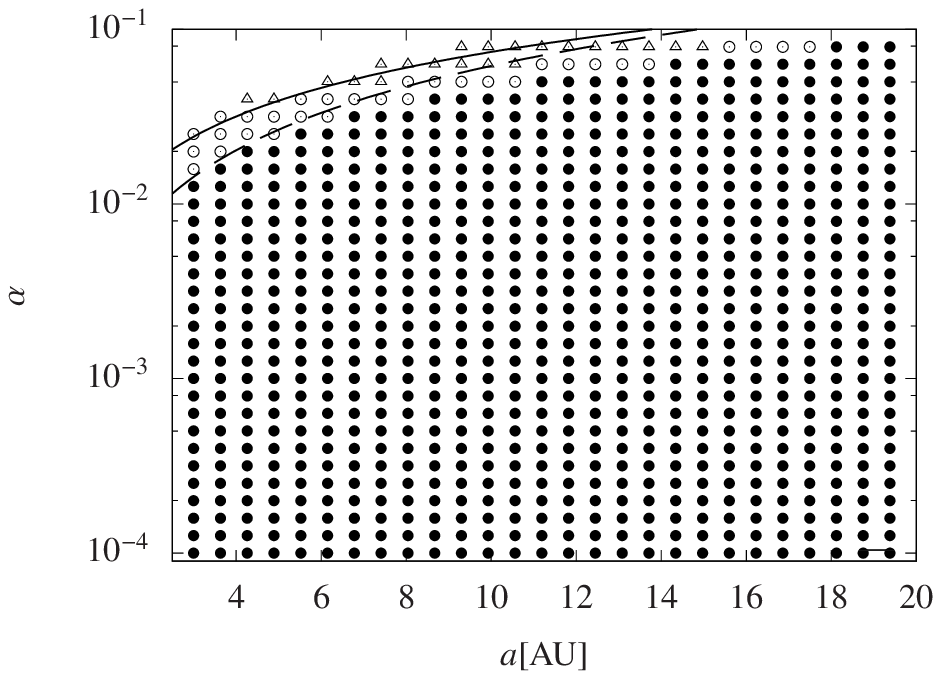}
\end{center}
\end{minipage} 
\begin{minipage}[b]{0.5\linewidth}
\begin{center}
(b) $\beta_\mathrm{t}=0.5$
\plotone{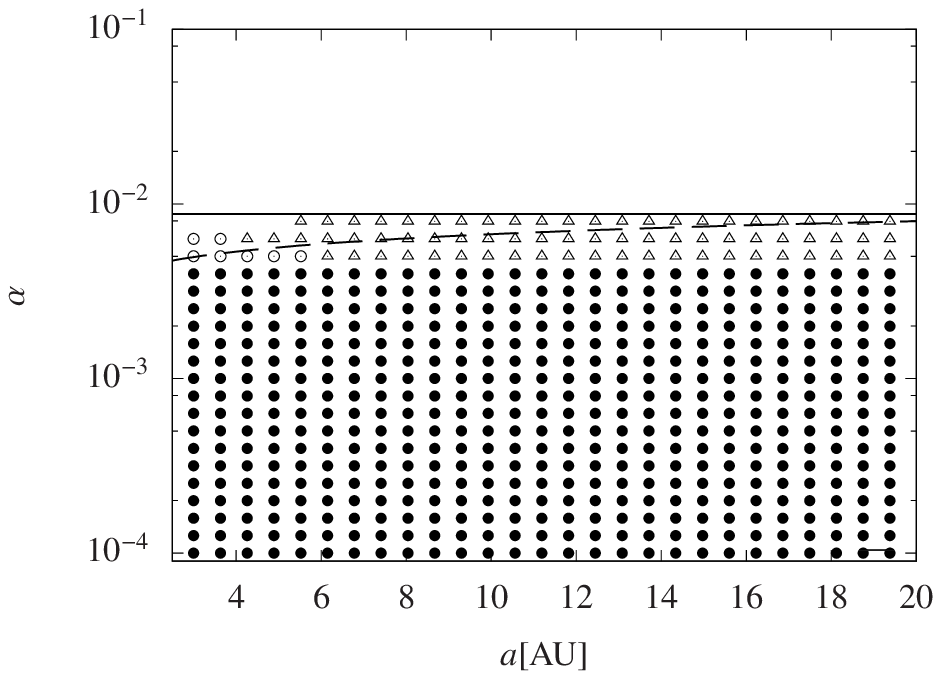}
\end{center}
\end{minipage} 
\begin{minipage}[b]{0.5\linewidth}
\begin{center}
(c) $\beta_\mathrm{g}=1.0$
\plotone{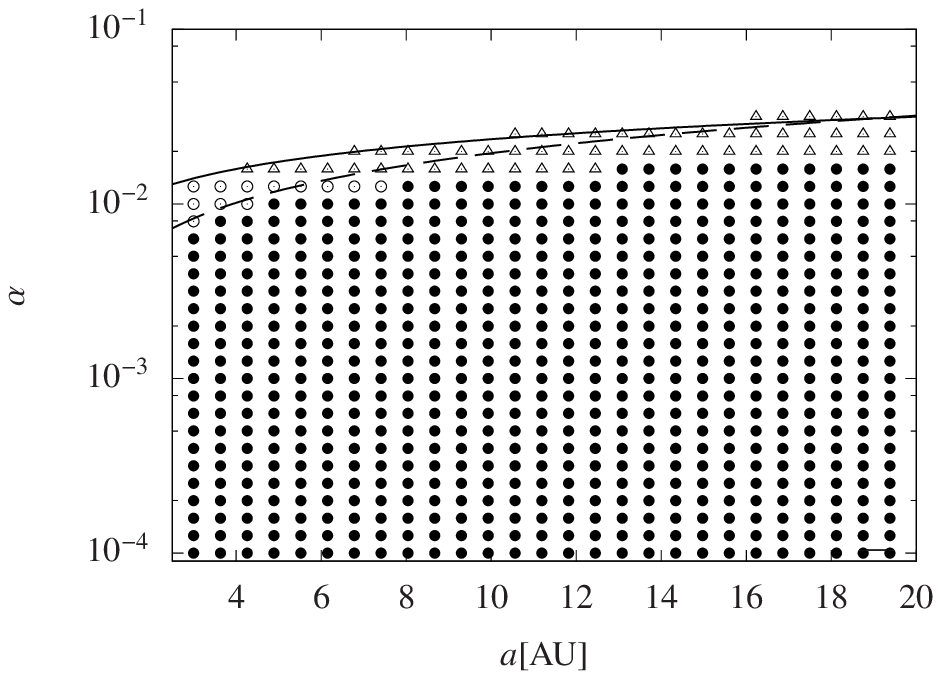}
\end{center}
\end{minipage} 
\begin{minipage}[b]{0.5\linewidth}
\begin{center}
(d) $\beta_\mathrm{t}=1.0$
\plotone{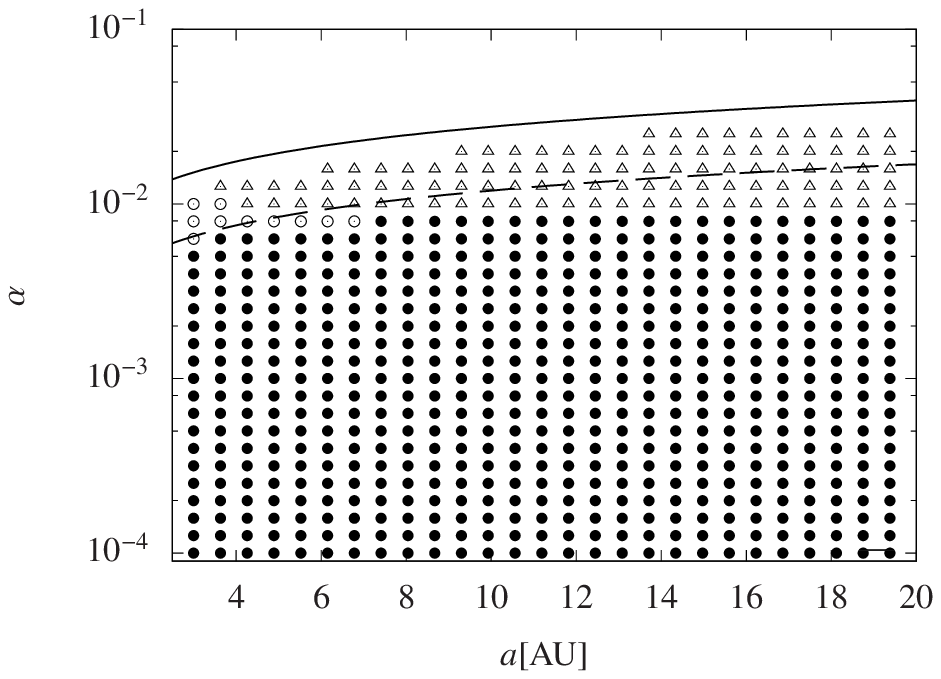}
\end{center}
\end{minipage} 
\begin{minipage}[b]{0.5\linewidth}
\begin{center}
(d) $\beta_\mathrm{g}=1.5$
\plotone{./gi_cond_a_alpha.eps}
\end{center}
\end{minipage} 
\begin{minipage}[b]{0.5\linewidth}
\begin{center}
(d) $\beta_\mathrm{t}=1.5$
\plotone{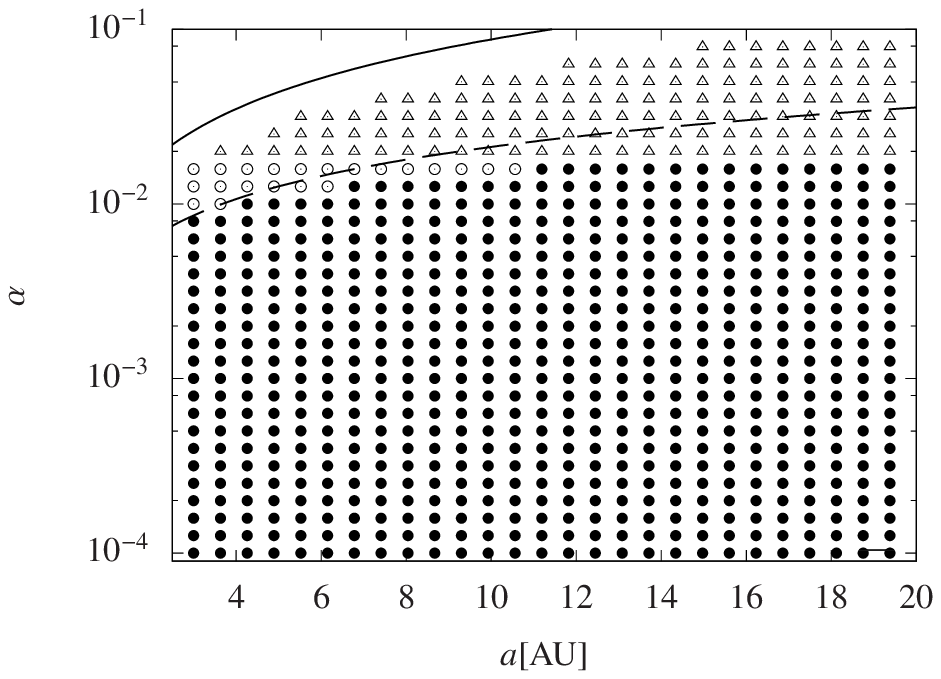}
\end{center}
\end{minipage} 
\caption{
Critical viscous parameter $\alpha$ for the GI versus $a$ for
(a) $\beta_\mathrm{g} = 0.5$, (b) $\beta_\mathrm{t} = 0.5$,
(c) $\beta_\mathrm{g} = 1.0$, (d) $\beta_\mathrm{t} = 1.0$,
(e) $\beta_\mathrm{g} = 1.5$, and (f) $\beta_\mathrm{t} = 1.5$.
The other parameters are the same as those in the fiducial model.
The symbols are the same as those in Figure \ref{fig:critical_alpha}.
}
\label{fig:critical_alpha2}
\end{figure}

Figure \ref{fig:critical_alpha2} shows the influence of the power-law
 indices $\beta_\mathrm{g}$ and $\beta_\mathrm{t}$. 
In the fiducial model, the dependence on $a$ is weak.
 However, if we adopt the general power-law indices, the dependence on
 $a$ can be strong.
As discussed in Section \ref{sec:cond}, the dependence on $a$ vanishes
 only if $\beta_\mathrm{g}-\beta_\mathrm{t} = 1$. 
As $\beta_\mathrm{g}$ increases, the dependence on $a$ becomes weak,
 while as $\beta_\mathrm{t}$ increases, the dependence on $a$ becomes strong.
This behavior is consistent with Equation (\ref{eq:cond}).

\subsection{Effect of Nonequilibrium Random Velocity \label{sec:noneq}}

In this section, we clarify the effect of the nonequilibrium random velocity on the onset of the GI. 
In the previous section, we assumed the equilibrium $\sigma_e$ and $\sigma_i$.
However, under some conditions, the relaxation time of $\sigma_e$ and $\sigma_i$ may be
 comparable to the growth time of dust aggregates. 
In this case, the equilibrium values change before $\sigma_e$ and
 $\sigma_i$ reach equilibrium.
Thus, the actual $\sigma_e$ and $\sigma_i$ may deviate from the
 equilibrium values.  

We will simultaneously consider the evolution of the mass and random velocity. 
The mass evolution equation is 
\begin{equation}
\frac{\mathrm{d}m_\mathrm{d}}{\mathrm{d}t} =
C_\mathrm{grow}
P_\mathrm{col} h^2 a^2 \Sigma_\mathrm{d} \Omega,
\label{eq:massevo}
\end{equation}
where $C_\mathrm{grow}$ is the sticking probability.
The perfect accretion corresponds to $C_\mathrm{grow}=1$.
For $C_\mathrm{grow} \ne 1$, we may need to change $C_\mathrm{col}$.
Strictly speaking, $C_\mathrm{col}$ depends on the energy dissipation rate on collisions,
such as the restitution coefficient and the friction coefficient on the dust aggregate surface.
For simplicity we assume $C_\mathrm{col}=1/2$ for any $C_\mathrm{grow}$.  
We solve the differential equations given as Equation (\ref{eq:evoe}), (\ref{eq:evoi}), and
 (\ref{eq:massevo}) for $\sigma_e$, $\sigma_i$, and
 $m_\mathrm{d}$, respectively.
For simplicity, we assume that dust aggregates always have
 the equilibrium internal density, $\rho_\mathrm{int} = \rho_\mathrm{eq}$.

\begin{figure}
\plotone{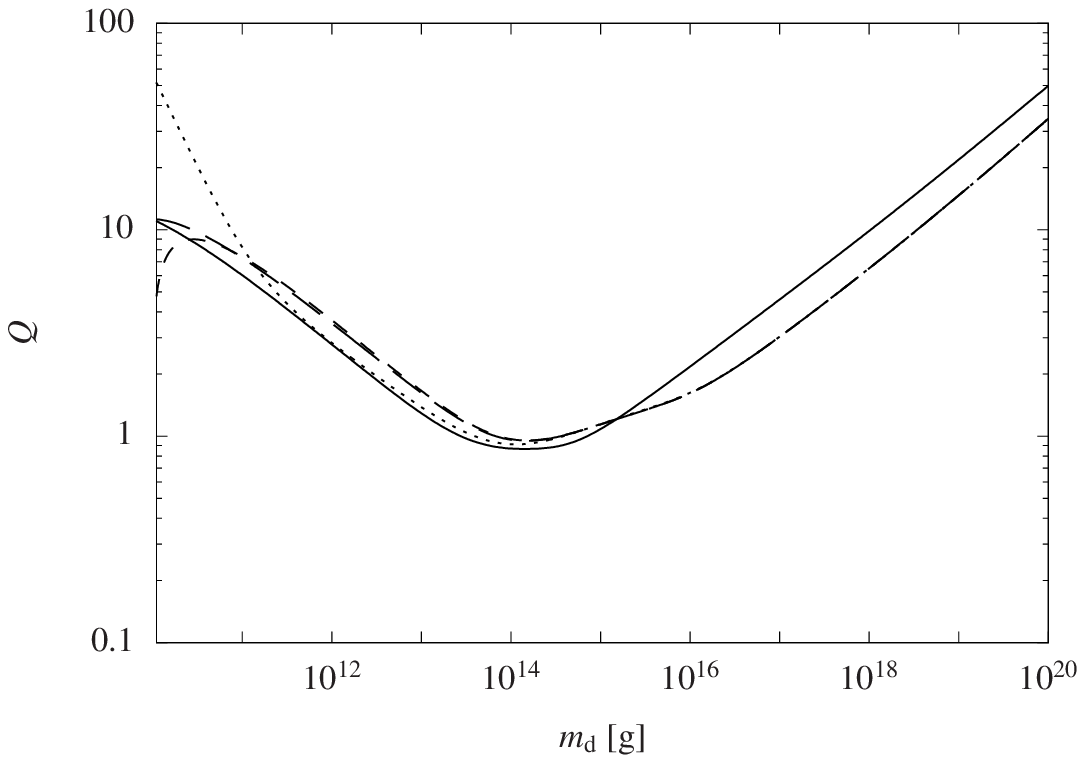}
\caption{
Evolution of $Q$ as the dust evolves in the fiducial model with the
 nonequilibrium and equilibrium (solid curve) random velocity. 
For the nonequilibrium model, the initial $\sigma_e$ and $\sigma_i$ are
$1/5$ (short-dashed), $1$ (dashed), and $5$ (dotted) times as large as the equilibrium values.
}
\label{fig:evolution}
\end{figure}

Figure \ref{fig:evolution} shows the evolution of $Q$
 for the fiducial model with an initial mass of
 $m = 10^{10} \mathrm{g}$ and initial equilibrium values for $\sigma_e$ and
 $\sigma_i$.
We assume the perfect accretion ($C_\mathrm{grow}=1$). As shown below, in the case of the perfect accretion, the difference between the equilibrium and nonequilibrium models is the most significant.
Toomre's $Q$ deviates slightly from the equilibrium value since the
 dust aggregates grow during the relaxation of the random velocity.
 Among the models shown in Figure \ref{fig:critical_alpha}, the deviation of the minimum $Q$ is $(9.7 \pm 13.2) \%$.
Thus, the nonequilibrium effect is not significant for the onset of the GI.
We also evaluated the influence of the initial values by considering values for $\sigma_e$ and $\sigma_i$ that were $5$
 times and $1/5$ times 
their equilibrium values.
As shown in Figure \ref{fig:evolution}, the difference in $Q$
 that stems from the initial values quickly vanishes as the system evolves.

\begin{figure}

\begin{minipage}[b]{0.518\linewidth}
\begin{center}
  \plotone{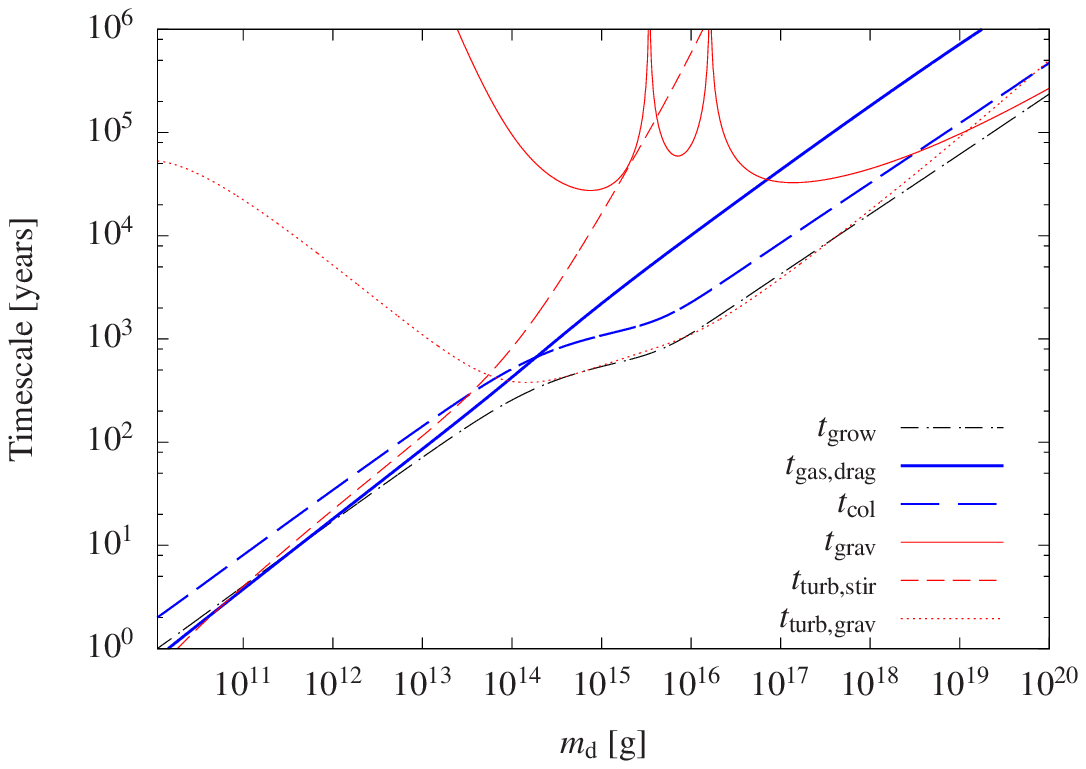}
\end{center}
\end{minipage}
\begin{minipage}[b]{0.482\linewidth}
\begin{center}
\plotone{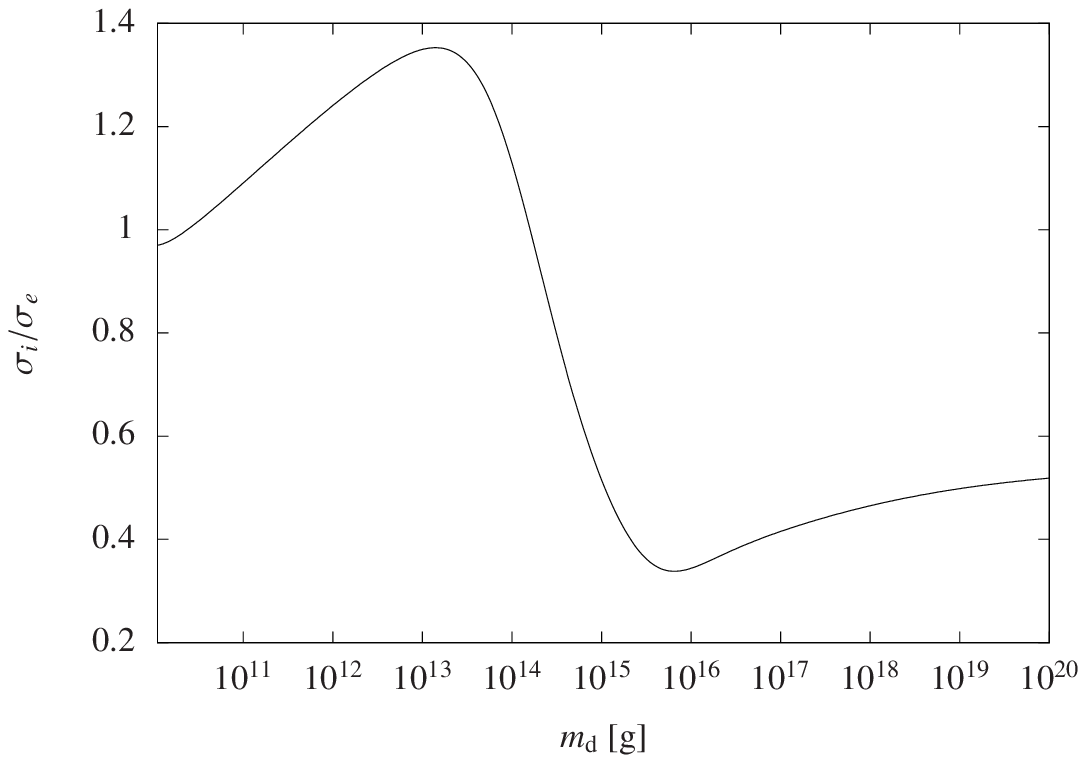}
\end{center}
\end{minipage}
  \caption{
    Various timescales (left) and the ratio of the inclination to the eccentricity (right) against the mass.  
    The timescales are $t_\mathrm{grow}$ (dashed-dotted), 
    $t_\mathrm{col}$ (thick solid),
    $t_\mathrm{gas,drag}$ (thick dashed),
    $t_\mathrm{grav}$ (thin solid),
    $t_\mathrm{turb,stir}$ (thin dashed), and
    $t_\mathrm{turb,grav}$ (thin dotted), respectively.
    The initial $\sigma_e$ and $\sigma_i$ are set as the equilibrium values.
  }
\label{fig:noneq_timescale}
\end{figure}

We compare the various relevant timescales.  The growth timescale is defined as
\begin{equation}
 t_\mathrm{grow} = m_\mathrm{d} / \frac{\mathrm{d}m_\mathrm{d}}{\mathrm{d}t},
\end{equation}
and the timescale of gravitational scattering is defined as
\begin{equation}
  t_\mathrm{grav} = \sigma_e^2 / \left| \left(\frac{\mathrm{d}\sigma_e^2}{\mathrm{d}t} \right)_\mathrm{grav} \right|.
\end{equation}
The other dynamical timescales $t_\mathrm{col}$,  $t_\mathrm{gas,drag}$,
$t_\mathrm{turb,stir}$, and $t_\mathrm{turb,grav}$ are defined in the same way.
The left panel of Figure \ref{fig:noneq_timescale} shows these timescales.
If the growth is sufficiently slow compared with the dynamical
timescales, the eccentricity reaches the equilibrium value. Figure
\ref{fig:noneq_timescale} shows that the growth timescale is comparable
with the dynamical timescales. Therefore the equilibrium eccentricity,
which depends on the mass, changes before the eccentricity converges
to the equilibrium value. However, since the growth timescale is not
very different from the dynamical timescales, the gap from the
equilibrium value is not so large.

Figure \ref{fig:noneq_timescale} shows two sharp peaks of $t_\mathrm{grav}$.
This is because $P_\mathrm{VS} $ is negative if $\sigma_i / \sigma_e$ is
small \citep{Ohtsuki2002}. In the relatively high-velocity cases,
gravitational scattering tends to realize $\sigma_i / \sigma_e \simeq 0.5$
\citep{Ida1990}. Therefore low $\sigma_i / \sigma_e$ leads to negative
$P_\mathrm{VS}$, while high $\sigma_i / \sigma_e$ leads to negative
$Q_\mathrm{VS}$. The right panel of Figure \ref{fig:noneq_timescale} shows
that $\sigma_i / \sigma_e$ is about $0.3$ around $m_\mathrm{d} \sim 10^{16} \,
\mathrm{g}$, then $P_\mathrm{VS}$ is negative. Around the points where
$P_\mathrm{VS} \simeq 0$, $t_\mathrm{grav}$ becomes very large.

\begin{figure}
\plotone{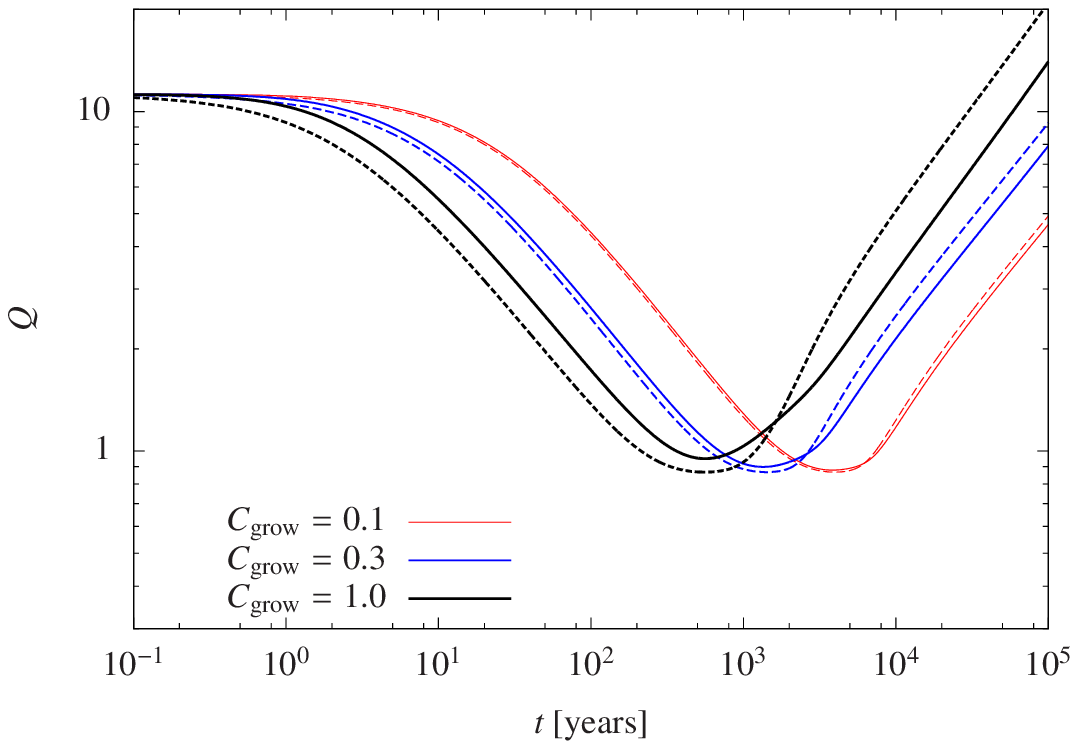}
  \caption{
      Time Evolution of $Q$ for $C_\mathrm{grow}=0.1$ (thin), $0.3$ (medium), and $1.0$ (thick) in the fiducial model.
      The solid curves corresponds to the nonequilibrium models and the dotted curves corresponds to the equilibrium models.
  }
\label{fig:cgrow}
\end{figure}

The effect of the imperfect accretion is shown in Figure \ref{fig:cgrow}.
The initial $\sigma_e$ and $\sigma_i$ are set as the equilibrium values.
The evolution timescale is the fastest for the perfect accretion model.
As $C_\mathrm{grow}$ decreases, the evolution timescale becomes longer,
which is inversely proportional to $C_\mathrm{grow}$.
The difference between the equilibrium and non-equilibrium models
becomes less with decreasing $C_\mathrm{grow}$.
This is because for smaller $C_\mathrm{grow}$, the growth time is
longer compared to the dynamical timescales and thus the eccentricity
can converge to the equilibrium value before the mass changes.

In Figures \ref{fig:critical_alpha} and \ref{fig:critical_alpha2},  we
examine the effect of the nonequilibrium velocity on the GI. If the nonequilibrium effect is considered, the
value of $\alpha$ necessary for crossing the GI region decreases slightly; however, the
difference is small. The nonequilibrium effect of the random velocity is thus
insignificant.
In summary, the above results justify the equilibrium random velocity model.

\subsection{Condition for Crossing the GI Region}

\begin{figure}
\plotone{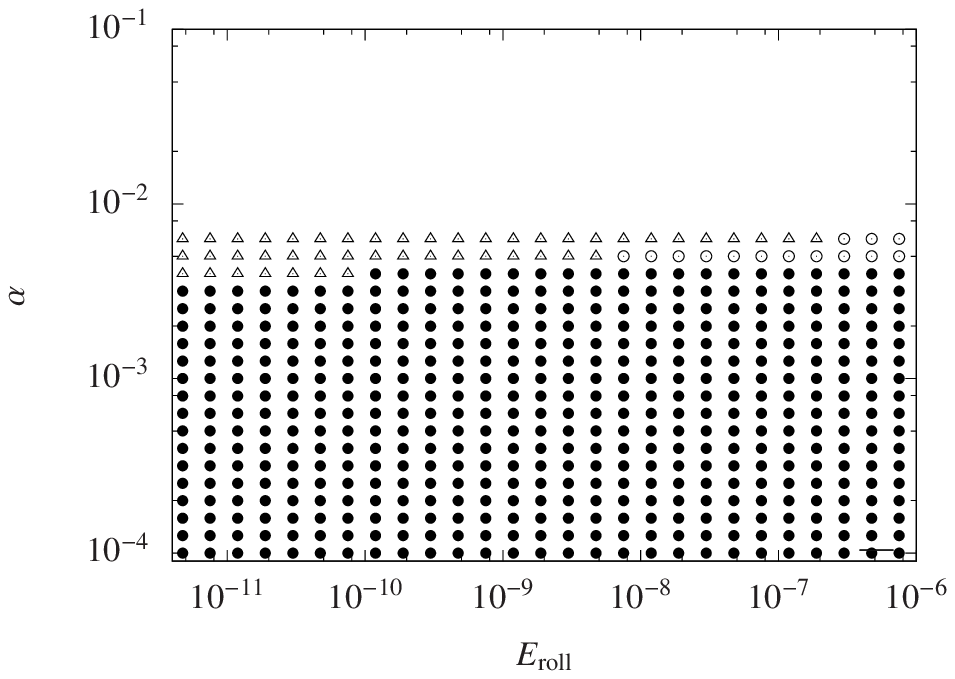}
\caption{
Same as Figure \ref{fig:critical_alpha} but versus $E_\mathrm{roll}$.
}
\label{fig:critical_alpha_eroll}
\end{figure}

\begin{figure}
\plotone{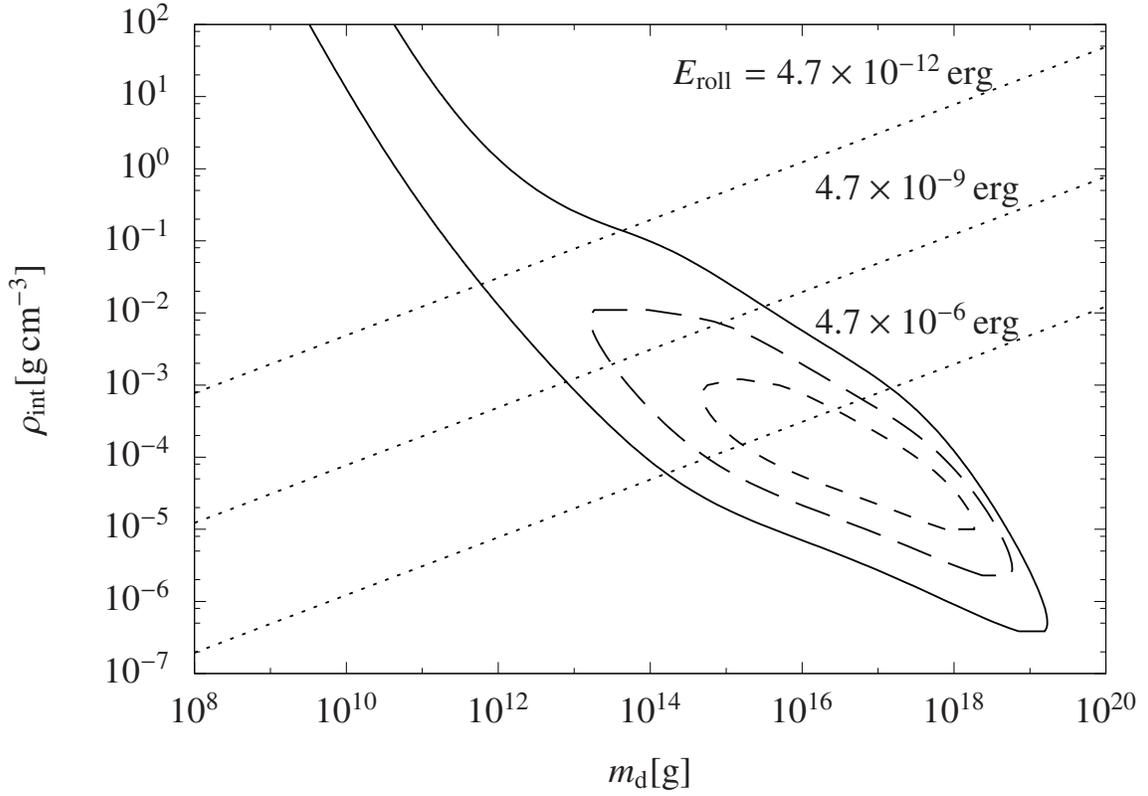}
\caption{
GI region for $\alpha = 2 \times 10^{-3}$ (solid), $4 \times 10^{-3}$
 (dashed), and $6 \times 10^{-3}$ (short dashed). 
The dotted lines show the dust evolution with
 $E_\mathrm{roll} = 4.7\times 10^{-12} \, \mathrm{erg}, 4.7 \times
 10^{-9}\, \mathrm{erg},$ and $4.7 \times 10^{-6}\, \mathrm{erg}$ from
 top to bottom, respectively. 
}
\label{fig:alpha_dp2}
\end{figure}

The dust evolution track is characterized by the monomer parameters
 $E_\mathrm{roll}$, $r_0$, and $\rho_0$. 
We first examine the effect of $E_\mathrm{roll}$.
In Figure \ref{fig:critical_alpha_eroll}, it can be seen that $E_\mathrm{roll}$ has little effect on the critical $\alpha$
 for crossing the GI region.
For small $E_\mathrm{roll}$, the critical $\alpha$ only slightly decreases. 
This is due to the structure of the GI region.
As shown in Figure \ref{fig:alpha_dp2}, the GI region stretches as
 $\alpha$ decreases.
In particular, the upper bound on the GI region increases rapidly with $\alpha$. 
For $\alpha = 2 \times 10^{-3}$, we can see the elongated GI region for
 $\rho_\mathrm{int}>0.1 \, \mathrm{g}\,\mathrm{cm}^{-3}$. 
If this region appears, the dust evolution inevitably crosses the GI region.  

\begin{figure}
\plotone{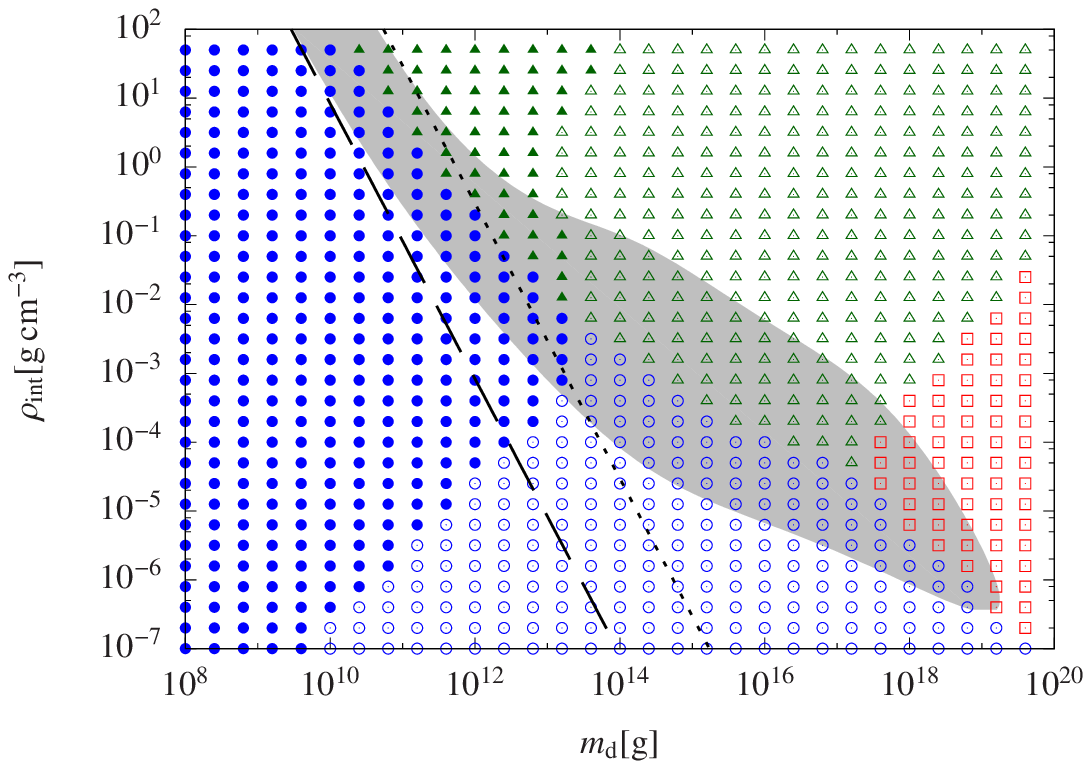}
\caption{
Main heating and cooling mechanisms for $\sigma_e$ with
 $\alpha = 2\times 10^{-3}$. 
The other parameters are the same as those of the fiducial model.
The symbols are the same as in Figure \ref{fig:source}. 
The dashed and dotted lines indicate the approximated boundaries of the
 elongated GI region described by Equations (\ref{eq:mlow,2}) and
 (\ref{eq:mhigh,2}), respectively.  
}
\label{fig:source2}
\end{figure}

To examine the condition for the existence of an elongated GI region, in Figure \ref{fig:source2},
 we show the main heating and cooling mechanisms for
 $\alpha = 2 \times 10^{-3}$.
The lower boundary of the elongated region is determined by the
 balance between turbulent stirring and gas drag.
   We calculate $(\mathrm{d} \sigma_e^2/\mathrm{d}t)_\mathrm{turb,stir}$ by the same way as that in Section \ref{sec:cond} except for $C_\mathrm{D}$.
Figure \ref{fig:various_ratio}e shows that Stokes drag $C_\mathrm{D} \simeq 24/\mathrm{Re}$ is a good approximation. Thus we adopt Stokes drag.
Using these approximations, we calculate $Q$. From $Q<2$, we obtain 
\begin{equation}
m_\mathrm{d} > m_\mathrm{low,2} =
2.78\times 10^{-2} \frac{
  \alpha^{3/2} 
  c_\mathrm{s}^{3/2} 
  \nu^{3/2}
  \tau_\mathrm{e}^{3/2} 
  M_*^{3/2}
  \Sigma_\mathrm{g}^{3/2}  
  }{
  Q_\mathrm{cr}^3
  a^{9/2}
  \rho_\mathrm{int}^{1/2}
  G^{3/2}
  \Sigma_\mathrm{d}^3 
  }. 
\label{eq:mlow,2}
\end{equation}
Similarly, the upper boundary of this region is determined by the balance between turbulent scattering and gas drag.
Thus, we obtain the upper boundary as
\begin{equation}
m_\mathrm{d} < m_\mathrm{high,2} =
5.62 \times 10^{3} \frac{
  Q_\mathrm{cr}^3
  \nu^{3/2} 
 \Sigma_\mathrm{d}^3
 }{
   \alpha^{3/2} 
   C_\mathrm{turb}^{3/2} 
   c_\mathrm{s}^{3/2}
  \rho_\mathrm{int}^{1/2} 
  \Sigma_\mathrm{g}^{3/2} 
}.
\label{eq:mhigh,2}
\end{equation}
The condition for the existence of an elongated region is
 $m_\mathrm{low,2} < m_\mathrm{high,2}$, from which we obtain the
 critical $\alpha$ for crossing the GI region:
\begin{equation}
  \alpha < \alpha_\mathrm{cr,2} =
58.7  \frac{
  Q_\mathrm{cr}^2 
  a^{2} 
  v_\mathrm{K}
 \Sigma_\mathrm{d}^2
 }{
   C_\mathrm{turb}^{1/2} 
   c_\mathrm{s} 
   \tau_\mathrm{e}^{1/2}
  M_* 
  \Sigma_\mathrm{g}
 }, 
\end{equation}
 which can be rewritten as
\begin{equation}
\alpha_\mathrm{cr,2} =
3.77 \times 10^{-3} 
f_\mathrm{g} \tau_\mathrm{e}^{-1/2} 
\left(\frac{\gamma}{0.018}\right)^2
\left(\frac{a}{1\, \mathrm{AU}} \right)^{\beta_\mathrm{t}/2-\beta_\mathrm{g}+3/2}
\left(\frac{C_\mathrm{turb}}{3.1 \times 10^{-2}} \right) ^{-1/2}
\left(\frac{M_*}{M_\odot}\right)^{-1/2}
\left(\frac{T_1}{120}\right)^{-1/2}. 
\label{eq:cond2}
\end{equation}
We show this condition in Figure \ref{fig:critical_alpha}.
If $\alpha < \alpha_\mathrm{cr,2}$, the dust evolution crosses the GI region.
Thus, we propose the following condition for the onset of the GI:
\begin{equation}
\alpha < \min(\alpha_\mathrm{cr,1}, \alpha_\mathrm{cr,2}).
\end{equation}

In deriving this condition, we did not consider the nonequilibrium
 effect of the random velocity. 
However, as shown in Figure \ref{fig:critical_alpha}, that effect is insignificant. 
For the outer disk ($a = 10\mbox{--}20\, \mathrm{AU}$), we need a safety factor, such as
 $\alpha < 0.5 \min(\alpha_\mathrm{cr,1}, \alpha_\mathrm{cr,2})$. 
 
In Paper I, we proposed a similar condition, and
its parameter dependence is exactly the same as that of
 $\alpha_\mathrm{cr,1}$. 
Other than when
 $\gamma$ is particularly large or particularly small, the difference between $\alpha_\mathrm{cr,1}$ and
 $\alpha_\mathrm{cr,2}$ is very small. 
Thus, the simple condition proposed
 in Paper I ($0.5 \times \alpha_\mathrm{cr,1}$) is also a good estimate of the condition for crossing the GI region in 
 most parameter regimes. 
 
\section{Summary and Discussion\label{sec:summary}}

We have investigated the stability of a disk that consists of porous
 icy dust aggregates in a turbulent gas disk. 
We calculated the random velocity of the aggregates, taking into
 account gravitational scattering, collisions, gas drag,
 turbulent stirring, and turbulent scattering.
In Paper I, we assumed an isotropic velocity dispersion and an
 equilibrium random velocity of the aggregates.
In this paper, we removed these assumptions.  
We separately calculated the evolution of the eccentricity and that of the inclination.
We found that under the evolution of dust by self-gravity compression, the GI is inevitable when the disk parameters are in a realistic range. 
In the minimum mass solar nebular model, the GI takes place when the
 turbulent viscosity parameter $\alpha$ is less than about
 $4 \times 10^{-3}$.  
We estimated the critical $\alpha$ and confirmed that it is in good agreement with numerical results.  
Thus, this estimate can be applied to general disk models. 

In this paper, we adopted the equal-mass dust aggregates for
simplicity. This assumption breaks down in some parameter ranges and
a size distribution develops. A wide or bimodal size distribution of
aggregates may alter the heating and cooling rates
quantitatively. Furthermore we need to include an additional dynamical
effect, dynamical friction between different sized aggregates. These
effects are out of the scope the present paper and should be
investigated separately.

Finally, we consider the post-GI evolution.
From linear analyses of the axisymmetric mode of the GI, the
 instability condition is $Q < 1$ \citep{Toomre1964, Goldreich1965a}. 
However, as pointed out by \cite{Michikoshi2007, Michikoshi2009,
 Michikoshi2010}, the axisymmetric mode does not appear in the dust disk.
This is because the nonaxisymmetric mode (gravitational wake) grows for
 $1 \lesssim Q \lesssim 2$, due to the swing amplification mechanism
 \citep{Goldreich1965, Julian1966, Toomre1981, Fuchs2001,
 Michikoshi2016, Michikoshi2016a}.
Initially, $Q$ of the porous-dust disk is sufficiently larger than $2$, and thus
 the disk is stable in any mode.
As the dust aggregate evolves due to self-gravity
 compression, $Q$ decreases gradually. 
For sufficiently small $\alpha$, $Q$ finally becomes less than $2$, and
gravitational wakes appear.
If the energy dissipation is effective and $Q$ decreases
 rapidly compared to the dynamical timescale, $Q$ may become
 less than unity, which leads to the development of the axisymmetric mode.
However, such a rapid decrease of $Q$ is unlikely.
The formation of a gravitational wake has also been confirmed by the
 hydrodynamics simulation of a dust layer \citep{Wakita2008}. 
\cite{Michikoshi2007} conducted $N$-body simulations that showed that the gravitational wakes fragment
 to form planetesimals. 
However, this fragmentation does not always take place.
For example, in a system where gas drag and inelastic collisions among
 particles are not effective, such as in a collisionless
 system, gravitational wakes develop due to the GI, but they do not
 fragment to form gravitationally bound objects
 \citep{Toomre1991,Fuchs2005,Michikoshi2014, Michikoshi2016}.
Therefore, energy dissipation in the wake is essential for
 planetesimal formation, and this suggests the existence of an additional
 condition for planetesimal formation.
The energy dissipation rate may be a key parameter, as it is in a gas
 disk, where if the cooling timescale is comparable to or shorter
 than the dynamical timescale, the disk fragments \citep{Gammie2001}. 
In our next paper, we will examine the post-GI evolution and
 scrutinize the necessary conditions for the formation of planetesimals.

The authors would like to thank the anonymous referee for useful comments.

\end{document}